\documentclass[fleqn,usenatbib]{mnras}

\usepackage{newtxtext,newtxmath}
\usepackage[T1]{fontenc}

\DeclareRobustCommand{\VAN}[3]{#2}
\let\VANthebibliography\thebibliography
\def\thebibliography{\DeclareRobustCommand{\VAN}[3]{##3}\VANthebibliography}

\usepackage{graphicx}	% Including figure files
\usepackage{amsmath}	% Advanced maths commands
\title[Ambiguous nuclear transients in ZTF]{A systematically-selected sample of luminous, long-duration, ambiguous nuclear transients}

\author[P. Wiseman et al.]{
\parbox{\textwidth}{
\Large
P. Wiseman,$^{1}$\thanks{E-mail: p.s.wiseman@soton.ac.uk (PW)}
R. D. Williams,$^{2}$
I. Arcavi,$^{3}$
L. Galbany,$^{4,5}$ %(0000-0002-1296-6887)
M. J. Graham,$^{6}$
S. H\"onig,$^{1}$
M. Newsome,$^{7,8}$
B. Subrayan,$^{9}$
M. Sullivan,$^{1}$
Y. Wang,$^{10}$ 
D. Ili\'c,$^{11,12}$ %$^{\orcidlink{0000-0002-1134-4015}}$
M. Nicholl,$^{13}$
S. Oates, $^{14}$
T. Petrushevska, $^{15}$
K. W. Smith,$^{13}$,
}
\vspace{0.04in}\\
$^{1}$School of Physics and Astronomy, University of Southampton, Southampton, SO17 1BJ, UK\\
$^{2}$Royal Observatory, University of Edinburgh, Edinburgh EH9 3HJ, UK\\
$^{3}$The School of Physics and Astronomy, Tel Aviv University, Tel Aviv 69978, Israel\\
$^{4}$Institute of Space Sciences (ICE-CSIC), Campus UAB, Carrer de Can Magrans, s/n, E-08193 Barcelona, Spain\\
$^{5}$Institut d'Estudis Espacials de Catalunya (IEEC), 08860 Castelldefels (Barcelona), Spain\\
$^{6}$California Institute of Technology, 1200 E. California Blvd, Pasadena, CA 91125, USA\\
$^{7}$Las Cumbres Observatory, 6740 Cortona Drive, Suite 102, Goleta, CA 93117-5575, USA\\
$^{8}$Department of Physics, University of California, Santa Barbara, CA 93106-9530, USA\\
$^{9}$Purdue University, Department of Physics and Astronomy, 525 Northwestern Ave, West Lafayette, IN 47907\\
$^{10}$Key Laboratory of Optical Astronomy, National Astronomical Observatories, Chinese Academy of Sciences, Beijing 100101, China\\
$^{11}$ Department of Astronomy, University of Belgrade - Faculty of Mathematics, Studentski trg 16, 11000, Belgrade, Serbia\\
$^{12}$Hamburger Sternwarte, Universitat Hamburg, Gojenbergsweg 112, D-21029 Hamburg, Germany \\
$^{13}$ Astrophysics Research Centre, School of
Mathematics and Physics, Queen's University Belfast, Belfast BT7 1NN, UK \\
$^{14}$Physics Department, Lancaster University, Bailrigg, Lancaster LA1 4YB, UK\\
$^{15}$University of Nova Gorica, Center for Astrophysics and Cosmology, Vipavska 11c, SI-5270 Ajdovščina, Slovenia\\
\\
}

\date{Accepted XXX. Received YYY; in original form ZZZ}

\pubyear{2024}

\begin{document}
\label{firstpage}
\pagerange{\pageref{firstpage}--\pageref{lastpage}}
\maketitle

\begin{abstract}
We present a search for luminous, long-duration ambiguous nuclear transients (ANTs) similar to the unprecedented discovery of the extreme, ambiguous event AT2021lwx with a $>150$\,d rise time and luminosity $10^{45.7}$\,erg\,s$^{-1}$. We use the Lasair transient broker to search Zwicky Transient Facility (ZTF) data for transients lasting more than one year and exhibiting smooth declines. Our search returns 59 events, seven of which we classify as ANTs assumed to be driven by accretion onto supermassive black holes. We propose the remaining 52 are stochastic variability from regular supermassive black hole accretion rather than distinct transients. We supplement the seven ANTs with three nuclear transients in ZTF that fail the light curve selection but have clear single flares and spectra that do not resemble typical AGN. All of these 11 ANTs have a mid-infrared flare from an assumed dust echo, implying the ubiquity of dust around the black holes giving rise to ANTs. No events are more luminous than AT2021lwx, but one (ZTF19aamrjar) has twice the duration and a higher integrated energy release. On the other extreme, ZTF20abodaps reaches a luminosity close to AT2021lwx with a rise time $<20$\,d and that fades smoothly in $>600$\,d. We define a portion of rise-time versus flare amplitude space that selects ANTs with $\sim50$ per cent purity against variable active galactic nuclei. We calculate a volumetric rate of $\gtrsim 3\times10^{-11}$\,Mpc$^{-1}$\,yr$^{-1}$, consistent with the events being caused by tidal disruptions of intermediate and high-mass stars.
\end{abstract}


%%%%%%%%%%%%%%%%%%%%%%%%%%%%%%%%%%%%%%%%%%%%%%%%%%

%%%%%%%%%%%%%%%%% BODY OF PAPER %%%%%%%%%%%%%%%%%%

\section{Introduction}

Untargeted, large-area photometric surveys have led to the identification of several intriguing classes of astrophysical transients whose characteristic timescales lie outside the weeks-to-month duration occupied by conventional supernovae. The optical transient parameter space now extends to fast, luminous, blue optical transients (RETs/FBOTs; e.g. \citealt{Drout2014, Pursiainen2018,wiseman2020a,ho_search_2023}) and slowly evolving superluminous supernovae (SLSNe; e.g. \citealt{Quimby2011, Nicholl2017, Inserra_statistical_2018, Angus2019, chen_hydrogen-poor_2023}). The destruction of stars by black holes -- tidal disruption events \citep[TDEs,][]{Hills1975,rees_tidal_1988} -- have also entered the era of sample studies \citep{Arcavi2014, Leloudas2019, Gezari_tidal_2021, VanVelzen2021, Nicholl2022, Charalampopoulos2022, Yao2023a,hammerstein_final_2023}. One of the most extreme and energetic transient phenomena of all, however, are still restricted to a handful of heterogeneously selected events whose discoveries have been serendipitous, whose observational classification is uncertain, and whose physical origin is unknown. So-called ambiguous nuclear transients (ANTs; e.g. \citealt{Kankare2017}; the most luminous of which have been named extreme nuclear transients, ENTs, \citealt{hinkle_extreme_2024}) appear to be related to supermassive black holes (SMBHs), their observational properties straddling the boundaries between single-star TDEs and continuously accreting active galactic nuclei (AGN). ANTs are loosely defined as nuclear transients (i.e. those spatially coincident with the nuclei of galaxies) whose combination of light curves and spectra do not fit any of the TDE, AGN or SN classes. They typically are characterised by long-lived, luminous optical flares with smooth rises and power-law decays \citep{Graham_understanding_2017}.

 AGN are inherently variable across the entire electromagnetic spectrum: in the ultra-violet--optical--near-infrared (UVOIR) the variability is stochastic on timescales from seconds to years, and is typically limited to amplitudes $\lesssim 0.5$\,mag on these timescales \citep[e.g.][]{vanden_berk_ensemble_2004,macleod_modeling_2010,caplar_optical_2017,sheng_lsst_2022}. Their variations are often described by a stochastic process called a damped random walk (DRW; \citealt{kelly_are_2009}). The presence of short-term variability in the optical is challenging to the canonical thin, viscous accretion disk model \citep[e.g.][]{lawrence_quasar_2018,antonucci_old_2018} and indicates that the disk reprocesses higher frequency (i.e. X-ray) emission from a small central region \citep[e.g.][]{clavel_correlated_1992,mchardy_origin_2016}. Such reprocessing, however, fails to explain the increasing number of AGN showing large-amplitude changes on short timescales: it is estimated that up to 30-50 per cent of quasars show variability at the $\sim1$\, mag level over baselines of 15\,yr \citep[][]{rumbaugh_extreme_2018}, dubbed `extreme variability quasars' (EVQs). Another growing class of AGN is that showing distinct spectral changes on years -- decade timescales: the (dis)appearance of broad emission lines \citep[e.g.][]{LaMassa2015,macleod_systematic_2016}, sometimes temporally coincident with a change in flux. So-called `changing-look AGN' (CLAGNs) challenge theories of AGN emission mechanisms even more (for a review see \citealt{ricci_changing-look_2022}). Correlations between the continuum luminosity and broad emission line appearance indicate that changes in the accretion flow drive CLAGNs, also called changing {\it state} AGNs\footnote{Note that changing-look has also been used to describe those AGN which change X-ray properties, which can also be denoted changing {\it obscuration} AGN \citep{ricci_changing-look_2022}}. In one extreme case, a non-variable AGN appears to `switch on' its variability over the course of a few years \citep{ridley_time-varying_2023} while another changes its spectral state on a timescale of months \citep{trakhtenbrot_1es_2019}.

The distinction between ANTs and AGN flares, CLAGNs, and TDEs, is not clearly defined. Indeed, TDEs are not excluded from occurring in AGN. ANT light curves appear to fall into two categories. Some ANT light curves are very smoothly evolving, rising slowly and decline monotonically even slower, including AT2019brs \citep{Frederick2021}, Gaia16aaw/AT2016dbs and Gaia18cdj/AT2018fbb \citep{hinkle_extreme_2024}, and the most energetic transient ever discovered, AT2021lwx \citep{Wiseman2023a,subrayan_scary_2023}. This behaviour appears to be a slower version of the evolution of most TDEs. Other ANTs, although displaying a single overall flare, show variability superimposed over their long-term light curves, such as AT2019fdr \citep{Frederick2021,reusch_candidate_2022,pitik_is_2022}, ASASSN-15lh \citep{Leloudas2016}\footnote{Interpreted in \citet{Leloudas2016} and \citet{Kruehler2018} as a TDE, but by \citet{Dong2016} as a SLSN.}, ASASSN-17jz \citep{Holoien2022}, ASASSN-18jd \citep{neustadt_tde_2020}, PS16dtm \citep[e.g.][]{blanchard_ps16dtm_2017,Petrushevska2023}, AT2017bgt \citep{Trakhtenbrot2019} and Swift J2219510-484240 \citep{oates_swiftuvot_2024}. While their light curves are similar to TDEs, ANTs are spectroscopically more similar to AGN with high equivalent width emission lines (particularly from the hydrogen Balmer series), while TDE emission lines are much broader but weaker. A subset of ANTs show features associated with Bowen fluorescence such as \ion{He}{ii} and \ion{N}{iii}, indicative of a steep far-UV or soft X-ray source such as accretion disk, and are termed Bowen fluorescence flares (BFFs; \citealt{Trakhtenbrot2019}). Some events classified as TDEs also have signatures of Bowen fluorescence, complicating the distinction between TDEs and ANTs further but hinting that the emission mechanisms are similar, irrespective of the source of accretion \citep{Leloudas2019, blagorodnova_iptf16fnl_2017,VanVelzen2021}. \citet{Frederick2021} presented a small sample of ANTs in known narrow-line Seyfert 1 (NLSy1) AGN, a class of Seyfert galaxies accreting close to the Eddington limit. PS16dtm also occurred in a NLSy1 \citep{blanchard_ps16dtm_2017}, while PS1-10adi also showed similar features \citep{Kankare2017}, raising the possibility that ANTs preferentially occur in such systems. 

An extra piece of the TDE - ANT - AGN puzzle is provided by mid-infrared (MIR) observations that probe hot dust. MIR flares, which tend to lag the UVOIR, accompany many of the UVOIR-discovered ANTs \citep{hinkle_mid-infrared_2024,Petrushevska2023, Wiseman2023a,oates_swiftuvot_2024}, as well as existing alone without optical counterparts \citep{Jiang2021, wang_mid-infrared_2022}. MIR flares are observed less frequently in TDEs \citep{VanVelzen2016}, although the population of MIR flares without optical counterparts \citep{Jiang2021,masterson_new_2024} could be caused by fully obscured TDEs. The interpretation is that these MIR-loud ANTs occur in nuclei with circumnuclear dust, typically described as the AGN `torus'. Nevertheless, some ANTs show no sign of previous AGN activity and optical spectra distinct from typical AGN \citep[e.g.][]{oates_swiftuvot_2024}.  

The recent discovery of AT2021lwx as the most energetic transient ever discovered pushed the boundaries of our understanding of SMBH accretion \citep{Wiseman2023a,subrayan_scary_2023}. For the extreme luminosity ($7\times10^{45}$\,erg\,s$^{-1}$) and duration ($>600$\,d in the rest frame), and thus total energy release ($>10^{53}$\,erg), to be explained by tidal disruption requires an unlikely combination of a large black hole mass ($10^{8.3}\,M_{\odot}$) and a massive ($15\,M_{\odot}$) star. Meanwhile the lack of forbidden oxygen and of any pre-cursory activity or shorter term variability makes an AGN implausible. In this paper we perform a systematic search for ANTs similar to AT2021lwx. We seek to understand whether it belongs to a population with similar energetics. We perform a census of nuclear transients with the aim of setting a standard distinction between ANTs and extreme but `standard' variability of AGN. 

In Section \ref{sec:sample} we introduce the search for ANTs in the Zwicky Transient Facility data stream. In Section \ref{sec:results} we present the individual ANTs and their shared properties. In Section \ref{sec:SF} we compare the variability properties of ANTs and AGN. We discuss the implications on the nature of ANTs in Section \ref{sec:discussion}, and conclude in Section \ref{sec:conclusion}. Where relevant we assume a spatially flat $\Lambda$-cold dark matter ($\Lambda \rm{CDM}$) cosmology with $H_0=70$\,km\,s$^{-1}$\,Mpc$^{-1}$, and $\Omega_M=0.3$. Unless otherwise stated, uncertainties are presented at the $1\,\sigma$ level. Magnitudes are presented in the AB system \citep{Oke1983}.

\begin{table}
	\centering
	\caption{Sample selection for ZTF nuclear transient candidates.}
	\label{tab:sample_selection_NT}
	\begin{tabular}{lccr} % four columns, alignment for each
		\hline
		Selection & Number cut & Number remaining \\
		\hline
		Initial selection & - & 53164 \\
		$<50$ points & 33187 & 19977 \\
		$t_{\mathrm{max}}-t_{\mathrm{min}}<1$\,yr & 1439 & 18538 \\
            $\Delta m <1$ & 15276& 3262\\
  Linear fit & 9165 & 58\\
  Visual inspection & 51 & 7\\
            
		\hline

  \end{tabular}
\end{table}
\begin{table}
	\centering
	\caption{Sample selection for ZTF orphan transients.}
	\label{tab:sample_selection_orphan}
	\begin{tabular}{lcc}
  \hline
		Selection & Number cut & Number remaining \\
		\hline
		Initial selection & - & 4529 \\
		$<50$ points & 3923 & 606 \\
		$t_{\mathrm{max}}-t_{\mathrm{min}}<1$\,yr & 331 & 275 \\
            $\Delta m <1$ & 268& 7\\
  Linear fit & 6 & 1\\
  Visual inspection & 0 & 1\\
		\hline
	\end{tabular}
\end{table}

\begin{table*}
	\centering
	\caption{Sample properties for the 11 events passing our selection criteria, split into smoothly evolving (upper) and spectroscopic (lower) selections.}
	\label{tab:sample}
	\begin{tabular}{lllllll} % four columns, alignment for each
		\hline
		ZTF ID & RA & Dec. &IAU name & Crossmatches & Host $r$-mag & Redshift \\
		\hline
             & & & & Photometric ANTs\\
            \hline
                ZTF20abrbeie& 318.45173&	27.43066&   AT2021lwx&	ATLAS20bkdj, PS22iin & No detected host&0.9945  $^{\rm abc}$  \\
                ZTF19aamrjar& 272.33963 &	25.30919&	   -  &ATLAS19mmu & 18.3 & 0.697     $^{\rm d}$ \\
                %ZTF19aaozooc& 209.93427 & 1.46943 & AT2021hum&	ATLAS20bjgp & 19.9&1.097     $^{\rm c}$ & Unclear\\
                %ZTF20aauowyg& 253.17170&	21.48924&  -        &ATLAS21ojg & 18.8&0.6612     $^{\rm a}$ & Unclear\\
                ZTF20abodaps& 359.80815&	-17.69658& AT2020afep& ATLAS20vrw & 19.8&0.607    $^{\rm d}$ \\
		    ZTF18aczpgwm   & 33.79096    &	7.16453    &AT2019kn& ATLAS19bdfo, Gaia19abv  &19.1 & 0.4279    $^{\rm d}$  \\
                ZTF21abxowzx& 324.33304&	-10.753312&   AT2021yzu&	ATLAS21bjoi & 20.3& 0.419 $^{\rm e}$  \\
                %ZTF20aaqtncr& 251.36440&	6.29627& AT2021fez&	ATLAS20pzv, Gaia21bgs & 19.7&0.368      $^{\rm b}$ & Unclear \\
                ZTF19aailpwl   & 216.94340 &	29.51061&AT2019brs&ATLAS19fyh, Gaia19axp  & 19.3 & 0.3736    $^{\rm f}$  \\
                ZTF20abgxlut & 257.76525 & 6.736331& AT2020oio &Gaia20dvv, ATLAS20rmk&  20.4 & 0.247$^{\rm g}$ \\
                ZTF20aadesap  & 232.76545 & 53.40535 & AT2022fpx &ATLAS22kjn, Gaia22cwy, PS23bdt & 17.9& 0.073$^{\rm h}$ \\
                
                ZTF18acvvudh &117.7084916	& 1.358392278&AT2018lcp & & 18.6 & -  \\
                ZTF22aaaeons & 231.827862	& -8.532609&AT2022zg &Gaia22aft, ATLAS22ery & 18.5 & -\\
                
                \hline
                 & & & & Supplementary spectroscopic ANTs\\
                \hline
                ZTF20acvfraq& 349.72403&	-10.58489&  AT2020adpi&	ATLAS20bjzp, Gaia21aid & 19.7&0.26  $^{\rm i}$ \\
                ZTF19aatubsj& 257.27857&	26.85569& AT2019fdr&	ATLAS19lkd & 18   &0.2666    $^{\rm j,h}$  \\
                ZTF20aanxcpf& 15.16508&	39.70842& AT2021loi&	ATLAS21qje & 18   &0.083     $^{\rm k}$  \\
                
                %ZTF20abgxlut& 257.7652339&	6.736335755& AT2020oio &ATLAS20rmk & 20.57 & 0.257   $^{\rm h}$\\
                
		\hline
            
  \end{tabular}
  
  \footnotesize{a) \citet{Grayling2022}; b) \citet{subrayan_scary_2023}; c) \citet{Wiseman2023a}; d) This work; e) \citet{chu_ztf_2021-1}; f) \citet{Frederick2021}; g) \citet{terreran_transient_2020}; h) \citet{perez-fournon_sglf_2022};  i) \citet{chu_ztf_2021};
  j) \citet{chornock_transient_2019};  k) \citet{makrygianni_at_2023}
            %h  \citet{Terreran_transient_2020};
            }
\end{table*}

\begin{figure*}
    \centering
    \includegraphics[width=\textwidth]{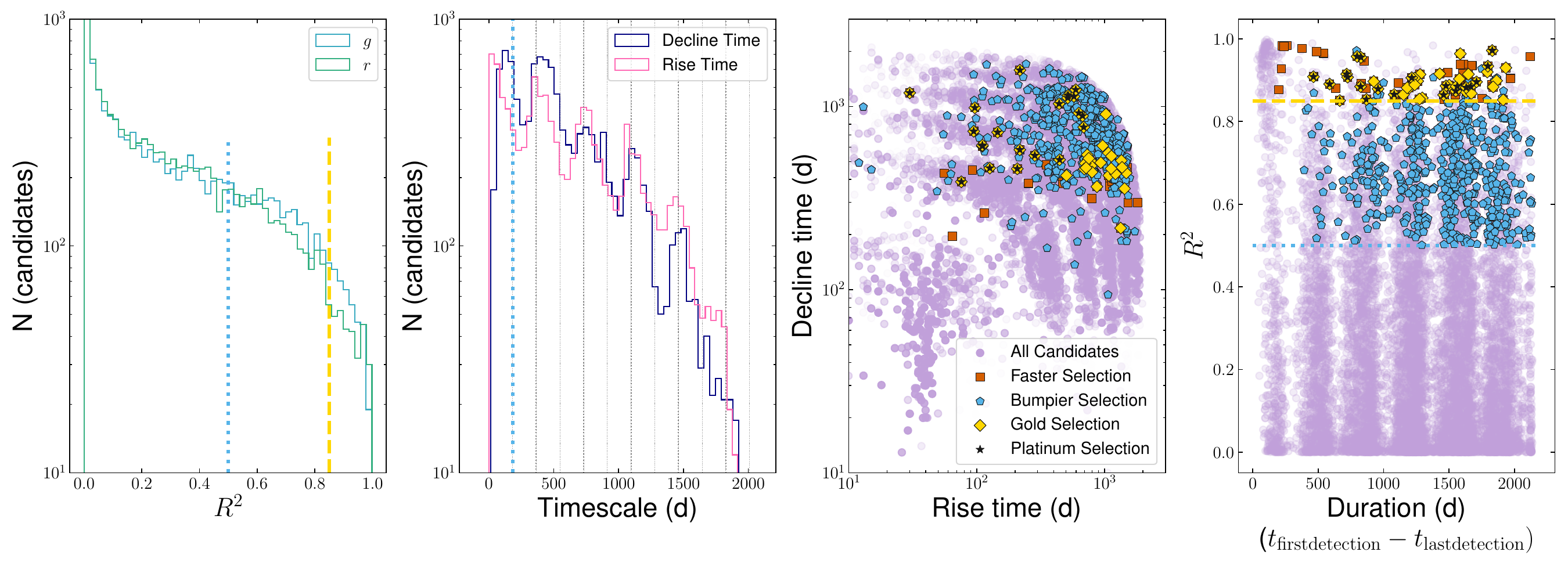}
    \caption{Light curve properties of nuclear transients in ZTF. In each panel the gold dashed and blue dotted lines indicate the threshold for the gold sample and bumpier samples, respectively. a) Histogram of the Pearson $R^2$ for the linear fit to light curve declines in the $g$ and $r$ bands; b) Histogram of the rise and decline timescales to the brightest point in the $r$-band light curve. 300 and 0.5 \,yr intervals are marked with vertical lines; c) $r$-band rise time versus decline time to/from the peak brightness; d) Pearson $R^2$ versus total $r$-band duration. }
    \label{fig:selection}
\end{figure*}
\section{Sample Selection}
\label{sec:sample}
We search for ANTs in the Zwicky Transient Facility (ZTF; \citealt{Bellm2019,Graham2019}).  ZTF is a wide-field, high-cadence survey that is extremely effective at detecting supernovae at low ($z\lesssim0.15$) redshift. \citet{Frederick2021} presented five ANTs serendipitously identified within the first two years of the ZTF survey. Of these, two (AT2019brs and AT2019fdr) are photometrically `AT2021lwx-like' with single UV-optical flares lasting over 1\,yr. The cadence, depth and now nearly 6-year baseline make ZTF the ideal survey to search for further ANTs. However, the selection of true nuclear transients compared to regular AGN variability or SNe is non-trivial \citep[e.g.][]{dgany_needle_2023}. In this section we describe the methods used to retrieve transients and identify ANTs in ZTF data.

\subsection{Filtering of slow, nuclear transients}
\label{subsec:lasair_search}
To search for ANTs in the ZTF data we use the \textsc{lasair}\footnote{\url{https://lasair-ztf.lsst.ac.uk/}} transient broker \citep{smith_lasair_2019} via its public application programming interface (API). The \textsc{lasair} API allows programmatic queries of the ZTF transient detection database via a number of criteria. We outline our initial selection criteria below.

\subsubsection{Nuclear transients}
ANTs are, by definition, located in galaxy nuclei. To select nuclear transients we use the \textsc{sherlock} sky context software \citep{Smith2020b} running in {\sc Lasair}. The ZTF search area is fully covered by the Panoramic Survey Telescope and Rapid Response System (Pan-STARRS1) science consortium surveys \citep{Chambers2016,Wainscoat2016}: \text{sherlock} uses the Pan-STARRS catalogues and probabilistic classifications of unresolved point sources \citep{tachibana_morphological_2018} to determine a host galaxy and the transient location within that host. We select all objects that are predicted as Nuclear Transients (i.e. transients consistent with a galactic nucleus) or AGN (transients consistent with a galactic nucleus that are listed in an AGN catalogue). Note that most transient searches exclude known AGN, but we allow them to avoid excluding bona-fide transient events associated with actively accreting SMBHs. We also include all objects predicted as SNe (transients not entirely consistent with a galaxy nucleus) with an angular separation of $<0.4\arcsec$ from the catalogued centre of a galaxy, based on the requirements used for the ZTF TDE samples \citep{VanVelzen2021,Yao2023a}. We verify the nuclear origin of any transients in our sample by validating that their mean position lies within $0.3\,\arcsec$ of the catalogued galaxy position, roughly equivalent to the ZTF positional uncertainty. Finally, we include ``Orphan" transients, which have no associated host galaxy in deeper survey imaging (as was the case for AT2021lwx). Including transient that appear hostless prevents the systematic exclusion of events at high-redshift, extremely dusty, or uncatalogued dwarf galaxies.

\subsubsection{Long-duration, high quality light curves}
The \textsc{lasair} API does not contain a parameter representing the duration of a light curve, as the queryable database only deals with 30-day alert packets\footnote{\url{https://zwickytransientfacility.github.io/ztf-avro-alert/}} rather than entire light curves. To filter out short-duration transients or those with poor coverage, we require the alert packet to have more than 10 detections (across both bands) of good quality and brighter than the reference (referred to as \texttt{ncandgp}$>10$).

\subsection{Photometric selection of ANTs}
\label{subsec:selection}
Roughly $50,000$ nuclear transient candidates and $\sim 4000$ hostless transient candidates pass our initial queries. We retrieve difference image photometry light curves via the \textsc{lasair} API of these events and make several further selections in order to identify ANT candidates. The number of events passing this selection are described in Tables \ref{tab:sample_selection_NT} and \ref{tab:sample_selection_orphan} for nuclear and hostless transients respectively. This selection applies to the ZTF $g$ and $r$ bands separately: an event failing either band will fail the overall cut.

We are aware that each of the following criteria, in particular the visual inspection, may introduce biases because we do not know what ANTs are. We are insensitive to events in off-nuclear black holes, as well as events with strong rebrightenings or variability during the decline, and events that are still ongoing (not yet declining) or did not have any detections on the rise. The nominal selection instead is aimed at selecting a high-purity sample of events similar to AT2021lwx. We subsequently test three alternative selections to verify the completeness achieved by the nominal selection.

\begin{enumerate}
    \item Light curve coverage: we require good quality light curve coverage, so events with fewer than 50 detections in each of the $g$ and $r$ bands are rejected. This selection removes objects with short-duration flares or noisy episodes.
    \item Duration: ANTs are long-lived transients, and we require that the most recent detection in at least one filter must be at least 300\,d after the first detection.
    \item Amplitude: AGN are naturally variable, but this variability is usually limited to a $\lesssim 0.5$\, mag over long timescales \citep[e.g.][]{macleod_modeling_2010}. To exclude most AGN variability we require events to show a brightening of at least $1$\, mag during the ZTF observing period. We make this selection on the difference between the brightest and faintest detection in either band.
    \item Single, smooth decline: Any remaining AGN and variable stars should show stochastic variability on timescales shorter than the typical decline of an ANT , a property we denote `bumpiness'. We define the opposite, `smoothness', as the lack of variability above the photometric noise on timescales shorter than the main lightcurve. Single transient events show power-law or exponential decays, which are linear in magnitude space. We fit the declines with a linear model and quantify the goodness-of-fit to the linear model with the Pearson $R^2$ parameter. We set an arbitrary threshold of $R^2>0.85$ for a decline to be accepted as smooth and linear. We vary this definition of `smoothness' below and determine its effect on the number of high-quality candidates to be minimal.
    
\end{enumerate}
The distribution of candidates in these parameter spaces is shown in Fig. \ref{fig:selection}.
After all the nominal selection criteria are applied, we are left with 58\footnote{59 separate candidate light curves passed, but we note that ZTF22aadesap and ZTF22aafuzjv are two identifiers for the same light curve, leaving 58 distinct transients. light curves of nuclear transients}  and one orphan (ZTF20abrbeie=AT2021lwx), which we denote the `gold selection'. Of the ANTs published by \citet{Frederick2021} we recover ZTF19aailpwl=AT2019brs (the other three are too `bumpy'). To explore ANTs with different light curve shapes, we subsequently trial a set of different selection criteria, which are described in Appendix \ref{sec:app_selection}. Briefly: we conduct a search for faster-declining events by relaxing the maximum duration to 180\,d (the `faster' sample); for less smooth events by relaxing the linear decline correlation coefficient to $R^2>0.5$ (the `bumpier' sample); and the decline time to be longer than the rise time (the `platinum' sample). The resulting samples and their ZTF light curve properties are available online\footnote{\url{https://github.com/wisemanp/ANTs-Nest}}. For the 58 gold events, we obtain forced photometry for the full ZTF survey duration via the ZTF forced photometry service \citep{Masci2019} for visual inspection.

\subsubsection{Visual inspection}

\begin{figure}
    \centering
    \includegraphics[width=.5\textwidth]{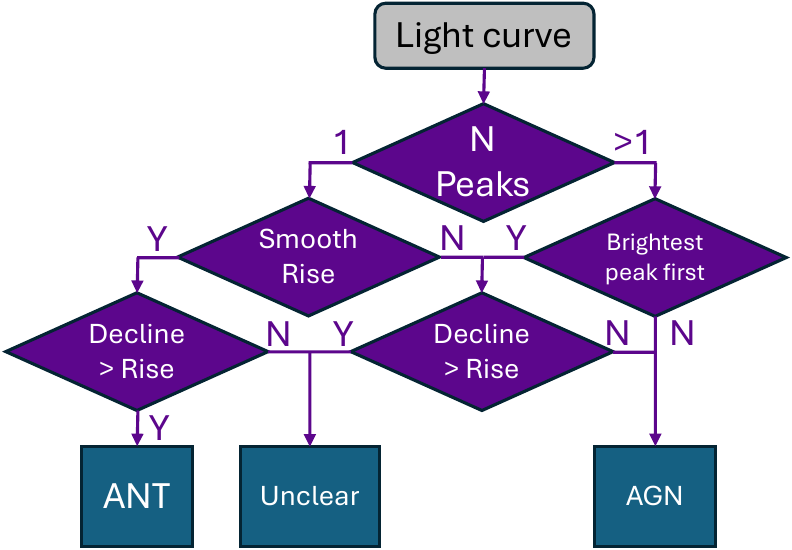}
    \caption{Flowchart used while visually inspecting the ANT candidates passing selection criteria. `Y' refers to a light curve passing criteria, while N is for failing. }
    \label{fig:flowchart}
\end{figure}

We visually inspect all 58 lightcurves and assign visual classifications of `Transient', `AGN', or `Unclear' according to the flowchart in Fig \ref{fig:flowchart}. These classifications are predominantly based on pre-and post-peak variability. In particular, the forced photometry allows us to view any variability that occurred before the candidate passed the alert threshold in ZTF. We classify 16 events as Transients, six as Unclear, and 36 as AGN.

\subsubsection{Spectroscopic history and follow-up}

For the analysis in this paper, we require spectroscopy for two reasons: firstly it is needed to remove non-ambiguous transients that were selected photometrically (i.e. supernovae and TDEs), and secondly in order to estimate the absolute luminosity, energetics, and the elemental composition of ANTs. 
Of the 58 `gold' light curves, 18 were reported as transients to the Transient Name Server (TNS)\footnote{\url{https://www.wis-tns.org/}}, and a handful have been publicly spectroscopically classified. Note that ANT is not a spectroscopic classification. Here we cross-match the candidates we photometrically classified as transients with the public classifications. In the cases where the spectra are clearly supernovae, we discard the events from our sample. We do not discard events with AGN-like spectra if the light curves are classified as Transient (these are, by definition, ANTs). We also check transients labelled as Unclear -- if they have AGN-like spectra we discard them as long-timescale AGN variability.

Of the 16 long-duration events we classified as transients via visual inspection, five are spectroscopically classified supernovae according to the public TNS classifications: ZTF19abpvbzf (SN~IIn; \citealt{brennan_epessto_2019}), ZTF22aanwibf (SN~II; \citealt{pellegrino_global_2022}). We also recovered the TDE ZTF21aaaokyp=AT2021axu \citep{hammerstein_ztf_2021,Yao2023a} which we discard as it is not ambiguous. 

The remaining events are detailed in Table \ref{tab:sample}. ZTF22aadesap=AT2022fpx \citep{perez-fournon_sglf_2022}: The spectra of ZTF22aadesap=AT2022fpx do not resemble the main classes of TDE defined by \citet{Arcavi2014, VanVelzen2021} and the object has recently been identified as an extreme coronal line emitter \citep{koljonen_extreme_2024}. The classification is thus ambiguous, so we include ZTF22aadesap=AT2022fpx as an ANT.

Seven of the remaining nine transients have been observed spectroscopically, four of which have public classifications while the other three are presented here for the first time. 
 ZTF19aailpwl=AT2019brs is an ANT \citep{Frederick2021}, and ZTF20abrbeie=AT2021lwx \citep{Wiseman2023a,subrayan_scary_2023} passes our selection by construction. ZTF21abxowzx=AT2021yzu and ZTF20abgxlut are spectroscopically classified as AGN at $z=0.419$ \citep{chu_ztf_2021-1} and $z=0.257$ respectively, but the light curves are clearly transient: in Section \ref{subsec:individuals} we classify these as ANTs. 

Further to public classifications, we obtained spectra for three that we photometrically and visually classified as Transients: ZTF18aczpgwm=AT2019kn, ZTF19aamrjar, and ZTF20abodaps=AT2020afep. These spectra are described in detail in Section \ref{subsec:individuals}. We also obtained spectra for three events which we labelled as Unclear: ZTF19aaozooc, ZTF20aaqtncr, ZTF20aauowyg. We discuss these events in Section \ref{subsec:uncertains}.
Finally, ZTF18acvvudh and ZTF22aaaeons have to our knowledge not been spectroscopically observed. They are therefore not considered for the remainder of this analysis, but we include them in Table \ref{tab:sample} for completeness.

After visual inspection and spectroscopic filtering, we retain seven ANTs which we term our `photometrically selected' sample: ZTF18aczpgwm=AT2019kn, ZTF19aailpwl=AT2019brs, ZTF19aamrjar, ZTF20abodaps=AT2020afep, ZTF20abrbeie=AT2021lwx, ZTF21abxowzx=AT2021yzu, ZTF22aadesap=AT2022fpx. We supplement these seven by searching for events within ZTF that fail our photometric selection but have similar spectroscopic features that do not resemble TDEs or AGN.

\subsection{Supplementary spectroscopic sample}
Our method of selecting ANTs with smooth power-law declines excludes genuine transient events with re-brightening episodes. To supplement this sample, we search for transients that have ANT-like light curves without the smoothness criteria. To do so we search the TNS for events that were observed spectroscopically and classified as AGN or type II superluminous supernovae (SLSNe-II) since ANT is not a well-defined spectroscopic class, and BFF is not a classification category in TNS. We inspect the light curves and spectra for ANT-like signatures: a nuclear location, a long-duration ($>300$\,d) light curve, and a spectrum with a blue continuum and strong Balmer emission lines. We find three ZTF-detected events that were not identified by our pipeline but that fit these criteria. Two of these have been published in the literature as ANTs (ZTF19aatubsj/AT2019fdr and ZTF20aanxcpf/AT2021loi), and one which has not (ZTF20acvfraq/AT2020adpi). We add these to our analysis sample. 

\begin{figure*}
    \centering
    \includegraphics[width=\textwidth]{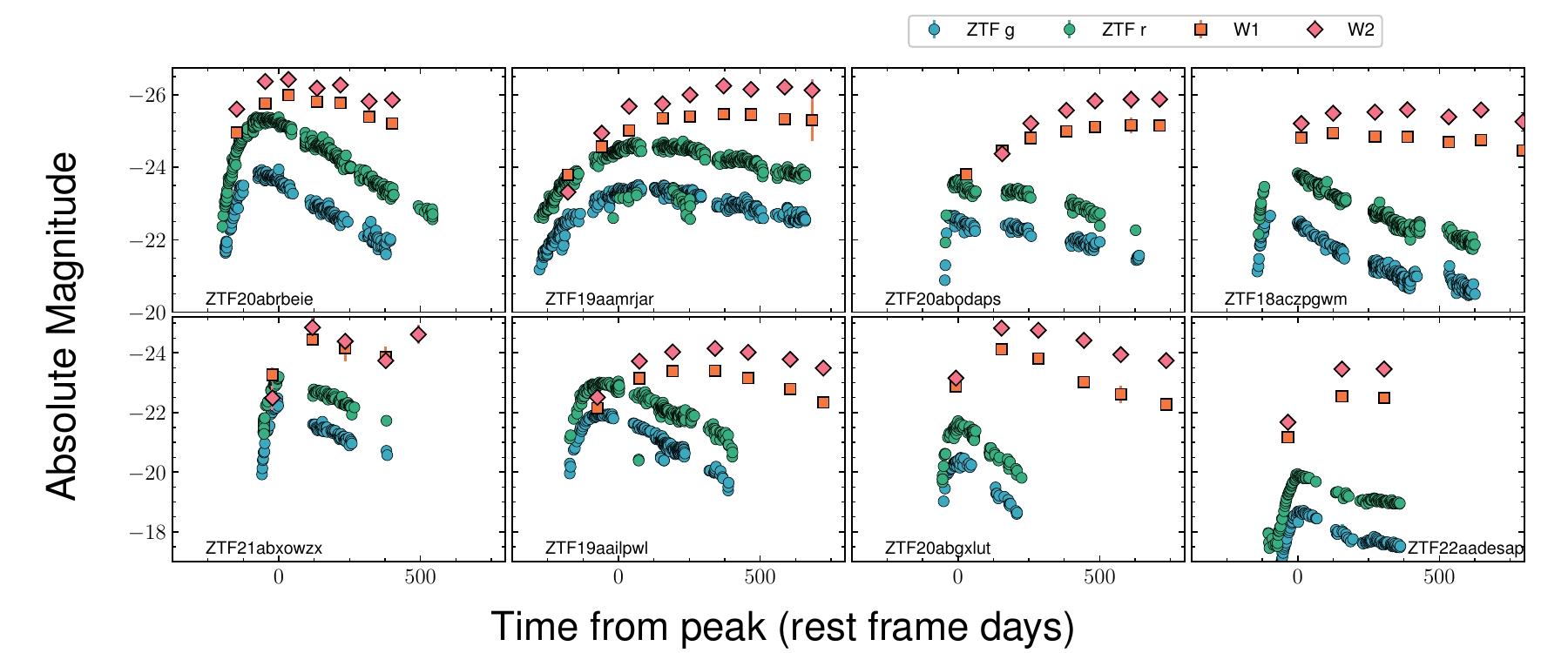}
    \caption{Optical and MIR light curves of the 8 photometrically selected ANTs passing `gold' cuts and with spectroscopic redshift. The $g$, $W1$ and $W2$ bands have been shifted arbitrarily for clarity.}
    \label{fig:lcs_gold}
\end{figure*}
\begin{figure*}
    \centering
    \includegraphics[width=0.75\textwidth]{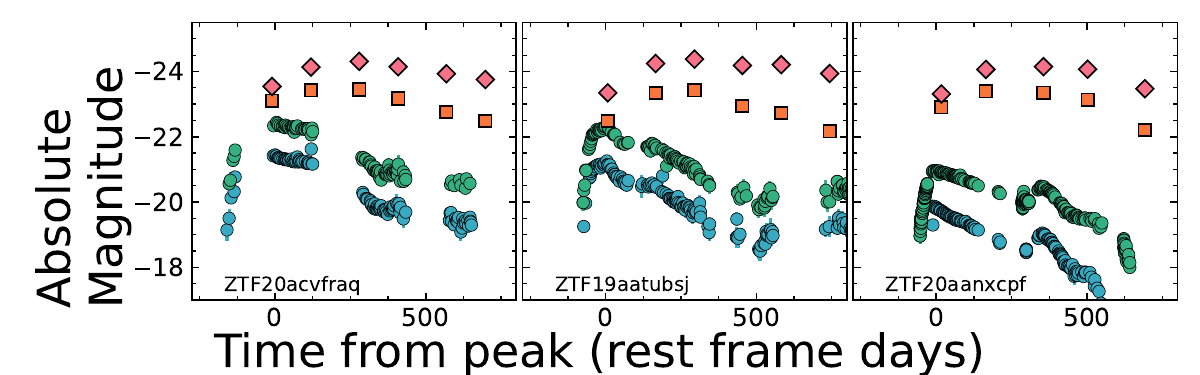}
    \caption{Optical and MIR light curves of the three spectroscopic ANTs with non-smooth light curves. The $g$, $W1$ and $W2$ bands have been shifted arbitrarily for clarity.}
    \label{fig:lcs_supp}
\end{figure*}

\begin{figure*}
    \centering
    \includegraphics[width=\textwidth]{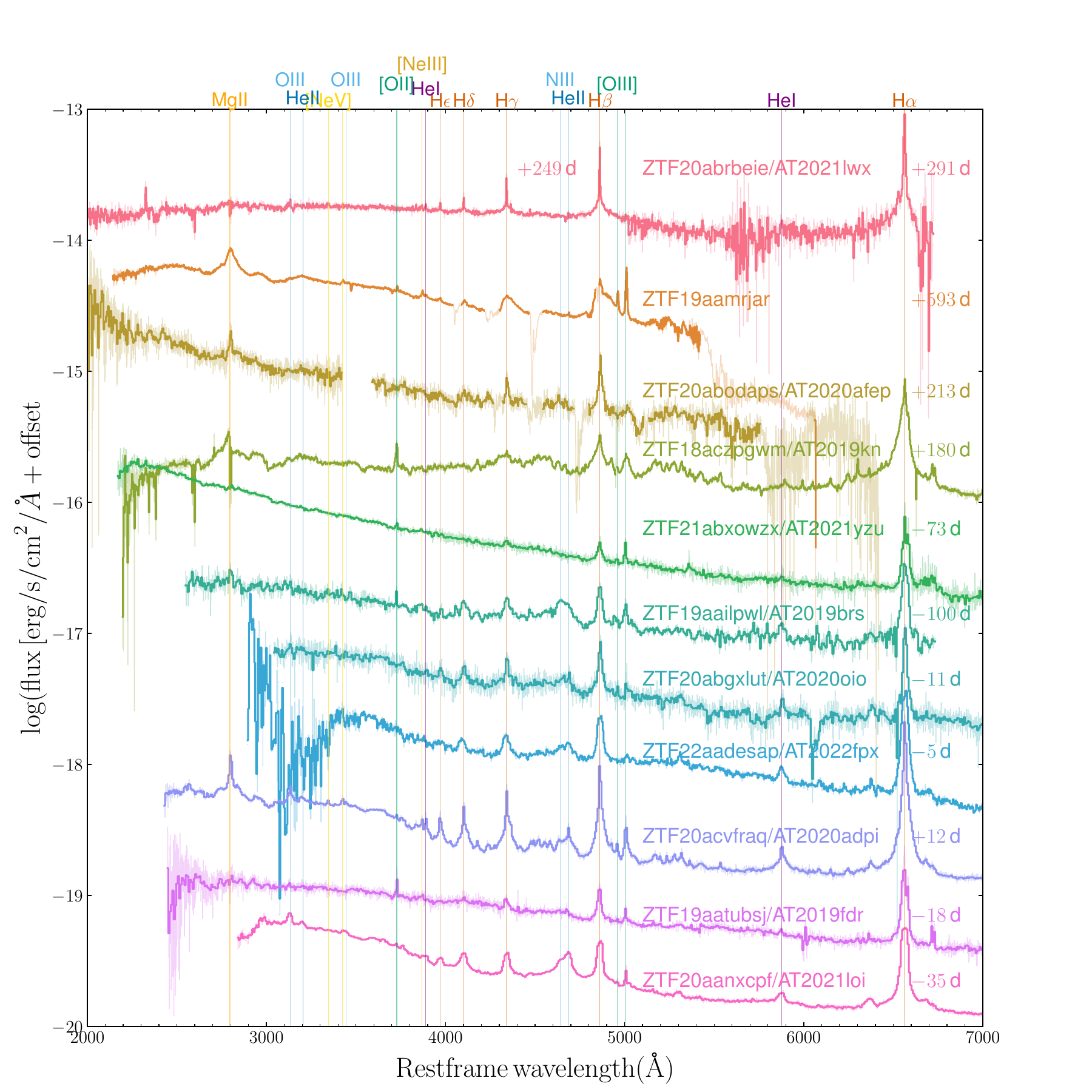}
    \caption{Rest frame UV-optical spectra of the 11 ANTs in the analysis sample. Spectra are logged before scaling to preserve colour. Data are presented at original resolution (light shading) as well binned to 2.5\,\AA \, resolution for aesthetics (dark line), except ZTF20aaqtncr/AT2021fez which is binned to 7.5\,\AA \, due to the low signal-to-noise.}
    \label{fig:specs_sample}
\end{figure*}

\begin{figure}
    \centering
    \includegraphics[width=.5\textwidth]{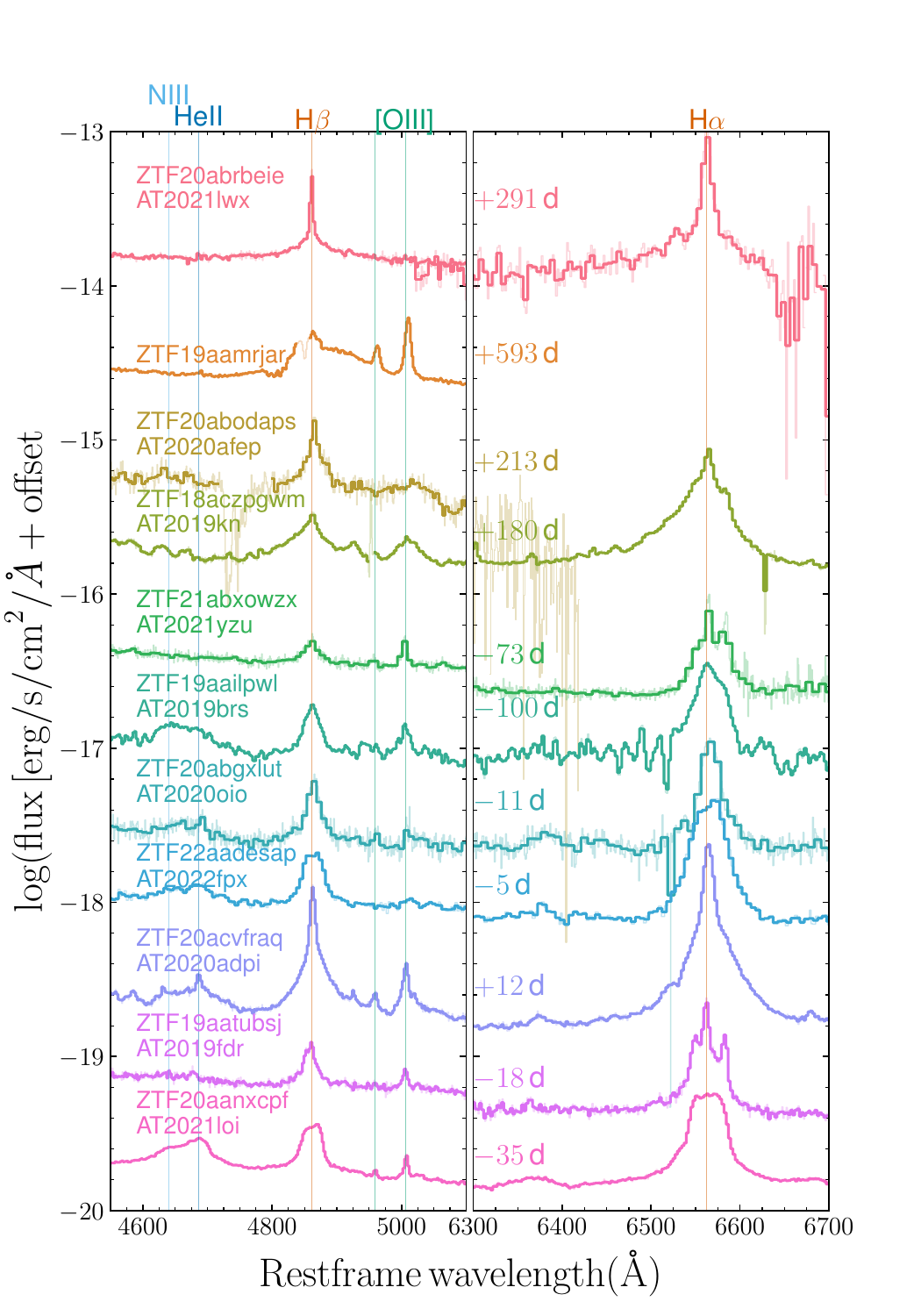}
    \caption{Left) The H$\beta$ [\ion{O}{iii}] complexes, also covering the \ion{He}{ii} and \ion{N}{iii} lines. Right) The H$\alpha$ line, where available. These spectra are identical to Fig. \ref{fig:specs_sample}, and have not been continuum subtracted.}
    \label{fig:specs_zoom}
\end{figure}

\section{A sample of ambiguous nuclear transients}
\label{sec:results}
Based on a photometric selection pipeline we identify 58 long-duration, smooth, high-amplitude events.  We classify 6 as unclear and 16 as transients, of which five are spectroscopically confirmed supernovae and one is a TDE. Of the remaining nine likely ANTs, seven have spectra that do not resemble supernovae or TDEs. We supplement these seven ANTs with three spectroscopically selected ANTs that have less smoothly evolving light curves. In this section we focus on the properties of these 11 selected ANTs, which we denote our `analysis' sample.

\subsection{Individual objects}
\label{subsec:individuals}
 We begin here by describing the individual events, before analysing the sample as a whole in Section \ref{subsec:sample_all}. For each object, we check for crossmatches in other surveys. For each survey we describe the data source and reduction methods below. In all cases, we correct for Galactic reddening using the dust maps of \citet{Schlafly2011} assuming a \citet{Fitzpatrick1999} extinction law.
 
 \subsubsection{ATLAS}
 We cross-match the ZTF transient with the database of the Asteroid Terrestrial-impact Last Alert System (ATLAS; \citealt{Tonry2018,Smith2020b}) from which we obtain forced difference imaging photometry from the ATLAS webserver \citep{shingles_release_2021}\footnote{\url{https://fallingstar-data.com/forcedphot/}}. Given the slow nature of the ANT light curves, to maximise signal-to-noise we combine ATLAS data into 20-day bins in each of the $c$ and $o$ filters.
 
 \subsubsection{Pan-STARRS}
 We obtain difference imaging photometry from the Panoramic Survey Telescope and Rapid Response System (PanSTARRS; \citealt{Wainscoat2016,Chambers2016}; a description of the transient filtering in current operation of the twin Pan-STARRS system is described in \citealt{smartt_gw190425_2024}).
 
 \subsubsection{Gaia}
 We obtain long-term, white-light \emph{Gaia $G$-band photometry from via Gaia Science Alerts \citep{hodgkin_gaia_2021}. These data are particularly useful for assessing past variability of a given source and are not used for parameter estimation, therefore we do not perform a host galaxy subtraction.}
 
 \subsubsection{WISE}
 We obtain mid-infrared photometry from the Wide-field Infrared Survey Explorer (WISE) spacecraft as part of the NEOWISE reactivation mission \citep{Mainzer2011} from the NASA/IPAC infrared science archive (IRSA\footnote{\url{https://irsa.ipac.caltech.edu/Missions/wise.html}}). We obtain catalogued photometry in the $W1$ ($3.4\,\mu m$) and $W2$ ($4.6\,\mu m$) bands for the full duration of the NEOWISE reactivation all sky survey (i.e. since 2013, five years before the start of ZTF). For each transient, we average the flux in each band over the period before the first optical detection and subtract this from post-optical-flare photometry.

 \subsubsection{VLASS}
For each transient we search the Karl G. Jansky Very Large Array Sky Survey (VLASS 2.1; \citealt{Lacy_karl_2020}) via the Canadian Astronomy Data Centre\footnote{\url{https://www.cadc-ccda.hia-iha.nrc-cnrc.gc.ca/en/vlass/}}.

\subsubsection{Swift}
We search the archive of the Neil Gehrels Swift Observatory \citep[{\it Swift;}][]{Gehrels2004}, and for five events obtained our own target of opportunity observations (PI: Wang). {\it Swift} X-ray data were taken in photon-counting mode with the X-ray Telescope (XRT; \citealt{Burrows2005}) and reduced with the tasks \textsc{xrtpipeline} and \textsc{xselect}. The source and background events were extracted using a circular region of 40\," and an annular ring with inner and outer radii of 60 and 110\,", respectively, both centred at the position of the source. The Ancillary Response Files were created with the task {\sc XRTMKARF} and the Response Matrix File (RMF), swxpc0to12s6\_20130101v014.rmf, was taken from the Calibration Data Base\footnote{\url{https://heasarc.gsfc.nasa.gov/docs/heasarc/caldb/swift}}. Targets were simultaneously observed with the Ultraviolet-Optical Telescope (UVOT; \citealt{Roming2005}) in the $uvw2$ (central wavelength, 1928 \AA), $uvm2$ (2246 \AA), $uvw1$ (2600 \AA), $u$ (3465 \AA), $b$ (4392 \AA), and $v$ (5468 \AA) filters. The task {\sc UVOTIMSUM} to sum all the exposures when more than one snapshot was included in each individual filter data and the task {\sc UVOTSOURCE} to extract magnitudes from aperture photometry. A circular region of 5\," centred at the target position was chosen for the source event, except for observations later than August 2023 after {\it Swift experienced a jitter, and thus we use 8\," -- this affects ZTF18aczpgwm, ZTF19aamrjar, ZTF20abodaps, ZTF21abxowzx}. Another region of 40\," located at a nearby position was used to estimate the background emission. We have not performed a host subtraction on the UVOT data. In cases where the UVOT data is used to constrain the black body SED, we verified that fits are consistent with and without the UVOT data.
 
 We search the for spectra in the following public archives: the European Southern Observatory (ESO) Science Portal \footnote{\url{https://archive.eso.org/scienceportal/home}}, the Keck Observatory Archive \footnote{\url{https://www2.keck.hawaii.edu/koa/public/koa.php}}, the Gemini Observatory Archive \footnote{\url{https://archive.gemini.edu/}}, the Gran Telescopio CANARIAS Public Archive \footnote{\url{https://gtc.sdc.cab.inta-csic.es/gtc/index.jsp}}. \\

Light curves for the analysis sample of 8 photometrically selected ANTs are shown in Fig. \ref{fig:lcs_gold} and the 3 spectroscopically selected are shown in Fig. \ref{fig:lcs_supp}. To convert to absolute magnitude for the figures we include a $1+z$ correction to preserve flux density, but have not $K$-corrected to the rest frame $g$ and $r$ bands. Spectra are displayed in Fig. \ref{fig:specs_sample}. Close-ups of the \ion{He}{ii} - [\ion{O}{iii}]-H$\beta$ region ($4500 - 5200$\,\AA) and the H$\alpha$ region ($6300-6700$\,\AA) is shown in Fig. (\ref{fig:specs_zoom}).
\vspace{1cm}
\subsubsection{Photometrically selected ANTs}
{\bf ZTF20abrbeie}/AT2021lwx/ATLAS20bkdj/PS22iin, $z=0.9945$. First reported by ZTF via the AMPEL broker. This event was analysed in \citet{Wiseman2023a} and \citet{subrayan_scary_2023}. The light curve is smooth and reaches $M_B\sim-25.7$, rising (in the rest frame) for over $>100$\,d and declining smoothly for over $450$\,d. It remains anomalous amongst the sample presented in this paper in having no detected host galaxy. The spectra displays broad and narrow Balmer emission lines, semi-forbidden carbon \ion{C}{ii}] and \ion{C}{iii}], and \ion{Mg}{ii}. There are no forbidden oxygen lines. A MIR flare is present with a $\sim 1$\,yr rise and subsequent decline. X-rays are detected from the source, consistent with a power-law and inconsistent with extrapolation of the UV-optical black body.\\ 

{\bf ZTF19aamrjar}/ATLAS19mmu, $z=0.697$. First reported by this work. 
ZTF19aamrjar has lasted just short of 1000\,d in the rest frame since the first detection by ZTF, with a rise and decline timescale greater than that of AT2021lwx\footnote{Since the initial sample selection and analysis was conducted, ZTF19aamrjar has shown a secondary rise, indicating either more persistent accretion consistent with the quasar-like spectrum or a rebrightening from the transient. Such rebrightenings further blur the line between ANTs and AGN.}. There is a MIR flare evolving slower than the optical, which has lasted 2\,yr and not yet reached its peak.  

This object was observed with the Las Cumbres Observatory \citep[LCO][]{brown_cumbres_2013} Floyds spectrograph mounted on the 2\,m Faulkes Telescope North in Haleakala, USA. The spectrum was reduced using the \texttt{floyds\_pipeline} \citep{valenti_first_2014}\footnote{\url{https://github.com/LCOGT/floyds\_pipeline}}. It was also observed with the Optical System for Imaging and low-Intermediate-Resolution Integrated Spectroscopy (OSIRIS; \citealt{cepa_osiris_2000}) on the 10.8\,m Gran Telescopio Canarias (GTC) at Observatorio del Roque de los Muchachos, La Palma, Spain. The spectrum was acquired in long-slit mode. The data were reduced via the \textsc{pypeit} package \citep{prochaska_pypeit_2020, prochaska_pypeitpypeit_2020}. The spectra show strong Balmer and \ion{Mg}{ii} emission at $z=0.697$. The Balmer lines (H$\beta$ -- H$\epsilon$) are very broad with full-width at half maximum (FWHM) of $\sim6200$\,km\,s$^{-1}$. The line profiles are asymmetric with red wings. [\ion{O}{ii}] and [\ion{O}{iii}] are strong and narrow. There is \ion{Fe}{ii} emission redwards of $5000$\,\AA. The spectrum is similar to highly luminous quasars with peculiar line profiles that have been identified as candidate recoiling or binary quasars \citep{Eracleous_large_2012,Shapovalova_first_2016}. 

ZTF19aamrjar was observed with {\it Swift} on MJD 60241.01 ($t_{\rm max}+634$\,d) for 2670\,s. The X-ray spectrum is a power-law with a photon index of  $1.8\pm0.9$ and a flux of $4.4\pm2.5\times 10^{-13}$\,erg\,cm$^{-2}$\,s$^{-1}$ in the 0.3-10\,keV range, corresponding to a luminosity of $9.5\pm5.4\times 10^{44}$\,erg
\,s$^{-1}$.\\

{\bf ZTF20abodaps}/AT2020afep/ATLAS20vrw, $z=0.607$. First reported by ePESSTO+. in this work. No previous variability in PS1 or PTF. The contrast between rise and decline timescales is extreme: the light curve rises to -24 mag in $\sim 30$\,d and declines constantly in 600\,d. Observed with LCO Floyds on FTN, and via the European Southern Observatory (ESO) as part of the extended Public ESO Spectroscopic Survey of Transient Objects (ePESSTO+; \citealt{Smartt2015}) using the ESO Faint Object Spectrograph and Camera (EFOSC2; \citealt{Buzzoni1984}) on the New Technology Telescope (NTT) at ESO La Silla observatory, Chile. The EFOSC2 spectrum was reduced using the PESSTO pipeline \citep{Smartt2015} v.3.0.1. The spectrum shows somewhat broadened Balmer lines (H$\beta$--H$\delta$) at $z=0.607$, along with \ion{Mg}{ii}. There is no detection of [\ion{O}{ii}] or [\ion{O}{iii}], similar to AT2021lwx. 

Similarly to ZTF19aamrjar, ZTF20abodaps/AT2020afep has a MIR flare evolving slower than the optical, which appears to have reached peak brightness 2\,yr after the optical peak. 

X-ray observations were obtained on MJD 60263.67 ($t_{\rm max}+738$\,d) with an exposure of 2367\,s. X-rays are marginally detected.We find an upper limit of $f_{0.3-10{\rm\,KeV}}<1.5\times10^{-13}$\,erg\,cm$^{-2}$\,s$^{-1}$ when assuming a power law with index $\gamma=1.8$ and a foreground absorption of $2.2\times10^{20}$\,cm$^{-2}$. This limit corresponds to a luminosity of $L_{0.3-10{\rm\,KeV}}<2.3\times10^{44}$\,erg\,s$^{-1}$.\\ 

 {\bf ZTF18aczpgwm}/AT2019kn/ATLAS19bdfo/Gaia19abv, $z=0.4279$. First reported by Gaia Science Alerts. The decline rate is very similar to AT2021lwx. A slight rebrightening occurred at +450\,d. 
 
 We obtain two publicly available spectra\footnote{Program IDs C252,C253; PI: Graham} from the Keck Low Resolution Imaging Spectrometer (LRIS; \citealt{Oke+1995}) at the W. M. Keck Observatory, Hawaii, USA. The spectra were reduced using the \textsc{LPIPE} pipeline \citep{perley_fully_2019} with default settings. The spectrum displays strong emission lines of the Balmer series and \ion{Mg}{ii} at $z=0.4279$, as well as broad forbidden lines of [\ion{O}{iii}] and narrow lines of [\ion{O}{ii}], \ion{O}{i}, and [\ion{S}{ii}]. The low-ionization lines (broad and narrow) show asymmetric profiles with a strong blue wing, indicating outflowing material. The presence of strong \ion{Fe}{ii} emission indicates a hot, ionizing UV continuum. In addition, high-ionization `coronal’ iron lines are present from [\ion{Fe}{vi}] and [\ion{Fe}{vii}]. The spectrum suggests an underlying AGN similar to a NLSy1. 
 
 The MIR flare rises for $\sim 1$\,yr, followed by a plateau lasting $>3$\, yr. 
 
 X-ray observations were obtained on MJDs 60262.82 and 60264.41 ($t_{\rm max}+1134$\,d, $+1136$\,d) with exposures of 1183\,s and 2580\,s respectively. We find an upper limit of $f_{0.3-10{\rm\,KeV}}<8\times10^{-14}$\,erg\,cm$^{-2}$\,s$^{-1}$ when assuming a power law with $\gamma=1.8$ and an absorption of $6\times10^{20}$\, cm$^{-2}$. This limit corresponds to a luminosity of $L_{0.3-10{\rm\,KeV}}<5.3\times10^{43}$\,erg\,s$^{-1}$.\\

{\bf ZTF21abxowzx}/AT2021yzu/ATLAS21bjhp, $z=0.419$. First reported by ZTF, classified as an AGN \citep{chu_ztf_2021}. A bona-fide flare, this event occurred in a galaxy with no previous variability and with pre-flare MIR colours (WISE $W1-W2$ versus $W3-W4$) not consistent with AGN activity.

We retrieve a Keck/LRIS spectrum from the Weizmann Interactive Supernova Data Repository (WISeREP; \citealt{Yaron2012})\footnote{ \url{https://www.wiserep.org/}}. The spectrum is the bluest of the sample and somewhat resembles a type I AGN \citep{vanden_berk_composite_2001}. It has broadened Balmer and \ion{Mg}{ii}, narrow [\ion{O}{ii}] and [\ion{O}{iii}], and hints of \ion{Fe}{ii}. 

There is a MIR flare with a timescale comparable to the optical light curve.

X-ray observations were obtained on MJDs 60243.54, 60248.44, and 60249.15 ($t_{\rm max}+495$\,d, $+499$\,d, $+500$\,d), with exposures of 2244\,s, 1138\,s, and 1518\,s respectively. We place an upper limit of $f_{0.3-10{\rm\,KeV}}<7.7\times10^{-14}$\,erg\,cm$^{-2}$\,s$^{-1}$ when assuming a power law with photon index 1.8 and an absorption of $3\times10^{20}$\, cm$^{-2}$. This limit corresponds to a luminosity of $L_{0.3-10{\rm\,KeV}}<4.9\times10^{43}$\,erg\,s$^{-1}$.\\

{\bf ZTF19aailpwl}/AT2019brs/Gaia19axp, $z=0.3736$. First reported by Gaia Science Alerts. Presented in the ANT sample of \citet{Frederick2021}. The light curve is similar to AT2021lwx, although less luminous ($M_B\sim24$) and with a faster ($\sim50$\,d) rise followed by a plateau and a similar gradient decline. 

The event occurred in a known NLSy1. The spectrum (retrieved from WISeREP) shows strong narrow Balmer emission along with \ion{He}{ii} and \ion{N}{iii} characteristic of Bowen fluorescence flares.

A MIR flare is present with a lag of $\sim 1$\,yr at peak.

There is extensive {\it Swift} coverage of this event, with observations grouped in two broad epochs: MJD 58550 -- 58850 ($t_{\rm max}-95$\,d to $+122$\,d), and MJD 59890 -- 60020 ($t_{\rm max}+880$\,d to $+975$\,d). There are no significant detections from each individual epoch, but grouping in the two above-defined bins reveals detections: assuming a power-law with $\gamma=1.8$ and Galactic absorption of $6\times10^{20}$\, cm$^{-2}$, the fluxes are $3.3 \pm 1.2 \times10^{-14}$\,erg\,cm$^{-2}$\,s$^{-1}$ and $7.7 \pm 4.9 \times10^{-14}$\,erg\,cm$^{-2}$\,s$^{-1}$ corresponding to luminosities of $1.6 \pm 0.6 \times10^{43}$\,erg\,s$^{-1}$ and $3.7 \pm 2.4 \times10^{43}$\,erg\,s$^{-1}$.\\

{\bf ZTF20abgxlut}/AT2020oio/Gaia20dvv/ATLAS20rmk, $z=0.257$. First reported by ZTF via the AMPEL broker. Initially classified as an AGN \citep{terreran_transient_2020}, there is no evidence for any variability in pre-existing data.

The transient spectrum (retrieved from WISeREP) is almost identical to ZTF19aailpwl/AT2019brs, except although the \ion{He}{ii}, \ion{N}{iii} as well as [\ion{O}{iii}] are all present, they are all weaker in ZTF20agbxlut. 

The pre-flare MIR colours (WISE $W1-W2$ versus $W3-W4$) are not consistent with AGN activity, and there is a clear MIR flare that lags the optical transient. There are no {\it Swift} X-ray observations of this event.\\

{\bf ZTF22aadesap}/AT2022fpx/ATLAS22kjn/Gaia22cwy/PS23bdt, $z=0.073$. First reported by ATLAS, this is the lowest luminosity event in our sample, showing an initially faster rise and decline followed by a quasi-plateau at later times.

ZTF22aadesap was classified as a TDE by \citet{perez-fournon_sglf_2022} but then identified as an ECLE by \citet{koljonen_extreme_2024}. While ECLEs have been observed in the late time stages of TDEs \citep{newsome_mapping_2024}, the peak spectrum of ZTF22aadesap/AT2022fpx does not resemble the common classes of TDE \citep[e.g.][]{Arcavi2014}.

X-ray observations do not show a signal during the optical peak, but a strong late-time flare is observed at the same time as the optical date plateau. The implications of these observations are discussed in \citet{koljonen_extreme_2024}.\\

\subsubsection{Spectroscopically selected ANTs}

{\bf ZTF20acvfraq}/AT2020adpi/ATLAS20bjzp/Gaia21aid, $z=0.26$. First reported by ZTF via the AMPEL broker. The rise time is uncertain due to gaps in the observed light curve, constrained to $40$\,d$<t_{\mathrm{rise}}<125$\,d in the rest frame. The initial decline is shallow, before a steep phase and then a second shallow phase.

We obtain a Keck/LRIS spectrum from WISeREP. The spectrum is similar to ZTf19aailpwl/AT2019brs: dominated by strong Balmer lines with broad (FWHM$=3000$\,km\,s$^{-1}$ and narrow components, along with \ion{He}{i} and \ion{Mg}{ii} with similar profiles. Similarly to ZTF19aailpwl/AT2019brs is the presence of \ion{He}{ii}, \ion{N}{iii}, and a particularly strong \ion{O}{iii} $\lambda\,3312$ line. These features indicate Bowen fluorescence. \ion{Fe}{ii} features can also be seen in the spectrum.

A MIR flare is present with a lag of $\sim 1$\,yr at peak. {\it Swift} observed the location of the transient over the period of a month around the optical peak. 

Swift XRT observed this event eight times between MJD 59411 and 59456 ($t_{\rm max}+20$\,d to $t_{\rm max}+56$\,d). In a total of $9.4$\,ks, no X-rays are detected. Assuming a power-law with photon index 2 this corresponds to an upper limit of $7.670\times10^{-14}\,\mathrm{erg\,s}^{-1}\,\mathrm{cm}^{-2}$, which equates to a luminosity limit $L_X < 1.6\times10^{43}\,\mathrm{erg\,s}^{-1}$.\\

{\bf ZTF19aatubsj}/AT2019fdr/ATLAS19lkd/PS19dar/Gaia19bsz, $z=0.2666$. First reported by ZTF via the AMPEL broker. Classified as a TDE in a NLSy1 \citep{Frederick2021,reusch_candidate_2022}, and although \citet{pitik_is_2022} classify it as a SLSN-IIn, we favour a SMBH-related interpretation.

The light curve is not smooth: the rise displays a `shoulder' before peaking after 70\,d, while after 70\,d of steep decline there is an 80\,d plateau before a shallower decline. After 500 days a 0.7\,mag rebrightening occurred and the transient has remained close to that level ever since. 

We retrieve the Keck/LRIS spectrum from WISeREP. The Balmer emission is the most prominent feature, and there is weak but clearly detected \ion{He{ii}.

There is a MIR flare with a lag at peak of $\sim 300$\,d. The spectrum has narrow + broad Balmer emission lines as well as \ion{Fe}{ii} emission. There is no clear detection of \ion{He}{ii} or Bowen fluorescence lines. A MIR flare is present with a lag of $\sim 1$\,yr at peak. 

X-ray observations are presented in detail by \citet{reusch_candidate_2022}. The event was observed by {\it Swift}-XRT with upper limits from observations in the first 315\,d reaching $f_{0.3-10{\rm\,KeV}}<1.4\times10^{-13}$\,erg\,cm$^{-2}$\,s$^{-1}$. However, there is a late time detection of soft X-rays (0.3-2.0\,KeV) from the eROSITA telescope \citep{predehl_erosita_2021} on board the Spectrum-Roentgen-Gamma spacecraft \citep{Sunyaev_SRG_2021} at MJD 59283.\\

{\bf ZTF20aanxcpf}/AT2021loi/ATLAS21qje, $z=0.083$. First reported by Gaia Science Alerts. This event was analysed by \citet{makrygianni_at_2023} as a Bowen fluorescence flare. It is the second lowest luminosity of our sample ($L_{\rm BB,max}=10^{44}$\,erg\,s$^{-1}$). It has a similar initial light curve shape to the majority of the analysis sample presented here, but displays a clear rebrightening after $\sim 400$\,d in the rest frame. 

The spectrum, from Keck/LRIS and obtained from WISeREP, is similar to ZTF19aailpwl/AT2019brs and ZTF20acvfraq/AT2020adpi in that it is dominated by Balmer lines with an unusual, flat-topped shape. There are strong emission lines from \ion{He}{ii}, \ion{N}{iii} and \ion{O}{iii}$\lambda\,3312$, indicating Bowen fluorescence and thus a very hot ionizing source. Any \ion{Fe}{ii} contribution is much weaker than in some of the Fe-strong events, but here the Balmer continuum is stronger. 

A MIR flare is present with a lag of $\sim 1$\,yr at peak.

\citet{makrygianni_at_2023} report X-ray upper limits from {\it Swift} XRT of $L(2-10\,\mathrm{keV}) < (2.4-7.8)\times10^{42}\,\mathrm{erg\,s}^{-1}$}.

\subsection{Unclear events}
\label{subsec:uncertains}

{\bf ZTF19aaozooc}/AT2021hum/ATLAS20bjat, $z=1.097$. The position of ZTF19aaozooc was targetted as part of the quasar survey \citep{richards_spectroscopic_2002} in the Sloan Digital Sky Survey \citep{York2000}. The spectrum, taken on MJD 52028 (19 years before the ZTF flare) and retrieved from SDSS data release 17 \citep{abdurrouf_seventeenth_2022}, is of a type I quasar with broad \ion{Mg}{ii}, \ion{C}{iii}] and \ion{C}{iv}. A further spectrum of ZTF19aaozooc was taken with Keck/LRIS around peak of the ZTF light curve (MJD 58788). The LRIS spectrum is qualitatively similar to the SDSS spectrum. We conclude that ZTF19aaozooc is a type I quasar with slow, high-amplitude variability.

{\bf ZTF20aaqtncr}/AT2021fez/ATLAS20pzv/Gaia21bgs, $z=0.368$. First reported by Gaia Science Alerts. The light curve has a smooth rise and decline of similar duration, although the ZTF data release photometry (without difference imaging) shows variability before the transient detection. A spectrum was obtained with LCO Floyds FTN. The spectrum shows strong narrow Balmer emission lines, strong [\ion{N}{ii}], and very strong [\ion{O}{iii}]. The [\ion{O}{iii}]/H$\beta$ ratio alone is enough to place the ionizing source as an AGN, which is confirmed by placing the line fluxes in the [\ion{O}{iii}]/H$\beta$ vs [\ion{N}{II}/H$\alpha$] Baldwin-Phillips-Terlevich (BPT) diagram \citep{Baldwin1981}: $\log(\ion{N}{ii}/\rm{H}\alpha) = 0.01$, $\log(\ion{O}{iii}/\rm{H}\beta=0.86$. A MIR flare follows the optical rise, but does not follow the decline. A lack of broad Balmer features suggests this event is a high amplitude variability from a type II (obscured) AGN, a population which tend to show smaller amplitude variability over large timescales \citep{de_cicco_structure_2022}. Nevertheless, such events could represent highly obscured examples of the more bonafide accretion transients and should be followed up with interest.\\

\begin{table*}
	\centering
	\caption{Properties of the transients presented in this sample. $\log L_{\mathrm{bol, max}}$, $\log T_{\mathrm{BB, max}}$, and $\log R_{\mathrm{BB, max}}$ are the luminosity, temperature and radius assuming a black body, fitted to the optical data at the peak of the optical light curve. Uncertainties are statistical only. $\log E_{\mathrm{rad}}$ is the radiated energy from the black body integrated over the observed light curve, with upper limits where the object is still visible. H$\beta$ widths are FWHM, and given for broad (b) and narrow (n) components.}
	\label{tab:bb_measurements}
	\begin{tabular}{llllllllll} % 
		\hline
		ZTF ID &IAU Name & $t_{\mathrm{rise}}$&$\log (L_{\mathrm{BB, max}})$   &$\log (T_{\mathrm{BB, max}})$   & $\log (R_{\mathrm{BB, max}})$&$\log (E_{\mathrm{rad, BB}})$   &H$\beta$ (b)&H$\beta$ (n) & $\log(L_{r,\mathrm{host}})$\\
             & & days&  erg\,s$^{-1}$ & K & cm & erg & km\,s$^{-1}$ & km\,s$^{-1}$ & erg\,s$^{-1}$\,\AA$^{-1}$\\
		\hline
                ZTF20abrbeie&  AT2021lwx & 198 & $45.9 \pm 0.1$          & $ 4.0 \pm 0.1 $ & $ 16.5 \pm 0.1 $& $>53.3  $ &2560$^{\rm a}$ & 250 & - \\
                ZTF19aamrjar&  -         & 377 & $ 45.6 \pm 0.4 $ & $ 4.2 \pm 0.1 $ & $ 16.0 \pm 0.1 $    &$>53.7$ & 6200$^{\rm b}$ &732 & 41.5\\
                %ZTF19aaozooc& AT2021hum  & 45.3$^{\rm c}$  & $^{\rm d}$     &\\
                %ZTF20aauowyg& -          & 45.4            & 3530           & \\
                ZTF20abodaps&  AT2020afep& 17 & $ 45.3 \pm 0.5 $ & $ 4.0 \pm 0.1 $ & $ 16.1 \pm 0.2 $& $ >53.1 $ & 1540            & & 40.7\\
		    ZTF18aczpgwm&  AT2019kn  &125 & $ 45.2 \pm 0.5 $&$ 4.0 \pm 0.03 $ & $ 16.2 \pm 0.05 $&$ 52.7 \pm 0.2 $ & 2831$^{\rm c}$  & 245 & 40.6\\
                ZTF21abxowzx&  AT2021yzu & 71 & $ 45.3 \pm 0.5 $&$ 4.3 \pm 0.03 $ & $ 15.7 \pm 0.03 $&$ 52.4 \pm 0.1 $ &1690            &194 & 40.2\\
                %ZTF20aaqtncr&  AT2021fez & 513 & $ 44.7 \pm 0.7^{\rm d}$  &$ 4.1 \pm 0.3^{\rm d} $ & $ 15.7 \pm 0.3^{\rm d} $&$ >52.5 $ &1320            &  & 40.3\\
                ZTF19aailpwl&  AT2019brs & 173 & $ 45.0 \pm 0.1 $ &$ 4.2 \pm 0.01 $ & $ 15.7 \pm 0.02 $&$ 52.5 \pm 0.1 $ &2650            &1000$^{\rm e}$ &40.4 \\
                ZTF20abgxlut& AT2020oio & 96 & $44.2 \pm 0.2$& $ 4.0 \pm 0.04 $ & $ 15.7 \pm 0.06 $&$ 51.3 \pm 0.4 $ & 3800 & 970 & 39.4 \\
                ZTF22aadesap& AT2022pfx &103  & $ 43.7 \pm 0.1 ^{\textrm{g}}$ &$ 4.1 \pm 0.02 $ & $ 15.2 \pm 0.04 $& $ 51.1 \pm 0.1 $ &  2100 & &39.4\\
                
                ZTF20acvfraq&  AT2020adpi& 120 & $ 44.6 \pm 0.2 $&$ 4.0 \pm 0.04 $ & $ 15.9 \pm 0.08 $&$ 52.3 \pm 0.1 $ & 3830            &537 & 39.9\\
                ZTF19aatubsj&  AT2019fdr & 77& $ 44.6 \pm 0.1 $&$ 4.0 \pm 0.03 $ & $ 15.9 \pm 0.05 $&$ 52.1 \pm 0.3 $ & 4100$^{\rm f}$  &370 & 40.1\\
                ZTF20aanxcpf&  AT2021loi & 59 &$ 44.0 \pm 0.1 $&$ 3.9 \pm 0.01 $ & $ 15.8 \pm 0.02 $&$ 51.4 \pm 0.1 $ &  2150            & & 39.5\\
                
                %ZTF20abgxlut& 257.7652339&	6.736335755& AT2020oio &ATLAS20rmk & 20.57 & 0.257   $^{\rm h}$\\
                
		\hline
            
  \end{tabular}
  
  \footnotesize{a) Revised from \citet{Wiseman2023a} based on higher quality data. An additional broad component with FWHM$\sim12000$\,km\,s$^{-1}$ is also compatible with the data; b) the broad line is highly asymmetric and affected by imperfect telluric subtraction; c) there may be an even broader component with FWHM$\sim 1.2\times10^4$\,km\,s$^{-1}$ but this may be caused by unmodelled \ion{Fe}{ii} emission; d) Formally unconstrained due to only two available filters constrain blackbody model;  e) consistent with the pre-flare measurement from an SDSS spectrum \citep{Frederick2021}; f) the significance of this component is dependent upon the continuum modelling; g) using $gri$ only as opposed to \citet{koljonen_extreme_2024} who add {\it Swift} UV data. }
\end{table*}

\begin{table*}
	\centering
	\caption{Dust properties of the transients presented in this sample. Luminosity, temperature and radius are measured at the MIR peak. A description of the various lags is provided in Section \ref{subsubsec:lags}. Lags have an uncertainty of $\pm90$\,d due to the NEOWISE cadence.}
	\label{tab:measured}
	\begin{tabular}{llllllllllll} % four columns, alignment for each
		\hline
		ZTF ID &IAU Name & $W1_{\mathrm{peak}}$ lag & $W2_{\mathrm{peak}}$ lag&$W1_{\mathrm{rise}}$ lag & $W2_{\mathrm{rise}}$ lag &$\log(L_{\mathrm{dust}})$&$\log(T_{\mathrm{dust}})$ & $R_{\mathrm{dust}}$ & $f_C$  &$\log(L_{\mathrm{bol, pre}})$ \\
             & & days & days & days & days & erg\,s$^{-1}$ & K & pc &  &erg\,s$^{-1}$ \\
		\hline
                ZTF20abrbeie&  AT2021lwx &119           & 119           &54  &54    & 44.63& 3.54&0.04& 0.06 & -\\
                ZTF19aamrjar&  -         &472        & 472        &25  & 25   & 44.55& 3.28&0.09& 0.09 & 44.9\\
                %ZTF19aaozooc& AT2021hum  &$>254$        & $>254$        &53  & 53   & 44.75& 3.30&0.10& 0.28\\
                %ZTF20aauowyg& -          &$>259$        & $>259$        &15  & 15   & 44.06& 3.29&0.05& 0.05\\
                ZTF20abodaps&  AT2020afep&$>492$         & $>492$         &220 & 220  & 44.69& 3.20&0.15& 0.16 & 44.6\\
		    ZTF18aczpgwm&  AT2019kn  & $>197^{\rm a}$& $>197^{\rm a}$ &77  & 77   & 44.42& 3.30&0.07& 0.17 & 44.4\\
                ZTF21abxowzx&  AT2021yzu &$>216^{\rm b}$& $100^{\rm b}$ &40  &$<181$& -&- &- &- & - \\
                %ZTF20aaqtncr&  AT2021fez &$>265^{\rm c}$ & $>265^{\rm c}$ &131 &  131 & 44.12& 3.19&0.09& 0.29 & 44.0 \\
                ZTF19aailpwl&  AT2019brs & 405          & 405          &116 & 116  & 44.41& 3.20&0.11& 0.20 &44.3 \\
                ZTF20abgxlut&  AT2020oio & 139          & 139          &43 & 43    & 44.13&3.16&0.09& 0.86 & - \\
                ZTF22aadesap& AT2022pfx &137            &286            &260 &69    &  43.02&3.02&0.05& 0.18& 43.5\\
                
                ZTF20acvfraq&  AT2020adpi&$305^{\rm c}$ & $305^{\rm c}$ &160 & 160  & 44.18& 3.12&0.12& 0.38 & - \\
                ZTF19aatubsj&  AT2019fdr &303           & 303           &82  & 82   & 44.00& 3.08&0.12& 0.25& 44.6\\
                ZTF20aanxcpf&  AT2021loi &173           & 364           &70  & 70   & 43.59& 3.14&0.06& 0.39 & 43.1\\

                %ZTF20abgxlut& 257.7652339&	6.736335755& AT2020oio &ATLAS20rmk & 20.57 & 0.257   $^{\rm h}$\\
                
		\hline
            
  \end{tabular}
  
  \footnotesize{ a) MIR light curve has a $\sim 600$\,d plateau; b) MIR detections are low significance and do not show a clear flare; c) time of optical and MIR peaks uncertain.}
\end{table*}
\subsection{Sample properties }
\label{subsec:sample_all}
In this section we present properties of the 11 selected ANTs measured from their light curves and spectra.

\subsubsection{Black body properties}
\label{subsubsec:blackbodys}
To estimate bolometric properties of the events, we model the observer-frame optical light curves as black bodies with evolving radii and temperatures using {\sc superbol} \citep{nicholl_superbol_2018}\footnote{\url{https://github.com/mnicholl/superbol}}. The resulting values are estimates that rely on several assumptions, namely that the emission is well described by a black body, in particular that it is spherically symmetric, and that thermal processes dominate over non-thermal emission. For three events, the presence of soft X-rays inconsistent with the UV-optical black body indicates either non-thermal emission or multiple black body emission regions. Likewise, the majority of the events show a delayed-onset mid-infrared (MIR) flare consistent with the dust echo mechanism (see Section \ref{subsubsec:lags}).
Black body measurements are presented in Table \ref{tab:bb_measurements}. The ANTs have peak black body luminosities in the range $44.7 \leq \log(L_{\mathrm{BB,max}}/\mathrm{erg\,s}^{-1}) \leq 45.85$, and are thus among the most luminous transient astrophysical phenomena known, in particular more luminous than canonical TDEs \citep{Yao2023a}. We integrate the black body luminosities over the full observed light curve\footnote{Note that some events are still ongoing, thus the energies are listed as lower limits} to obtain the total radiated energy $E_{\mathrm{rad, BB}}$. The inferred energies are ten times greater than the most energetic supernovae \citep[e.g.][]{Nicholl2020} and three times the most energetic TDEs \citep{Leloudas2016,Andreoni2022}, and are firmly in the $E_{\mathrm{rad}} > 10^{52}$\,erg regime of the heterogeneously compiled ANTs in the literature \citep[e.g.][]{Leloudas2016, Graham_understanding_2017, Kankare2017, oates_swiftuvot_2024}.

\subsubsection{Spectroscopic properties}
\label{subsubsec:spec_properties}

The ANT spectra are clearly distinct from those of TDEs, which are dominated by blue continua and low equivalent width broad emission lines - some show only H, while others show H+He, He alone, Bowen fluorescence lines and Fe, or are `featureless' \citep{Arcavi2014,Leloudas2019,VanVelzen2021, Charalampopoulos2022}. Our 11 ANTs are more AGN-like than TDE-like, consistent with the majority of ANTs in the literature. ANTs have narrower emission lines of higher equivalent width. Balmer lines of H are ubiquitous and have narrow and broad components, but the broad components are not as broad as those in TDEs: the mean H$\beta$ width from our 11 ANTs is $\sim 2900$\,km\,s$^{-1}$ (Table \ref{tab:bb_measurements}) compared to $>10^4$\,km\,s$^{-1}$ for TDEs \citep{Charalampopoulos2022}. Eight out of 11 ANTs have clear, narrow [\ion{O}{iii}]\,$\lambda5007$ which is a tracer of low-density ionized gas and present in both AGN and star-forming galaxy spectra \citep{vanden_berk_composite_2001,Osterbrock1989a}. It is lacking in the spectra of ZTF20abrbeie/AT2021lwx and ZTF20abodaps/AT2020afep. The ratio of this line to H$\beta$ can be used to estimate whether the ionizing source is an AGN or star formation \citep{Baldwin1981,Kewley_theoretical_2001,Kauffmann2003} while its absence indicates that neither AGN emission or star formation are strong in the hosts of these transients.
%\subsubsection{X-ray properties}
%\label{subsubsec:xrays}
\subsubsection{Mid-infrared lags}
\label{subsubsec:lags}
All 11events show changes in their MIR luminosity as measured in the WISE W1 ($3.3\mu$m) and W2 ($4.6\mu$m) bands. The light curve morphologies are varied, but all show a rise that is delayed from the optical flare. ZTF18aczpgwm, ZTF19aamrjar, ZTF20abodaps, ZTF20aaqtncr, and ZTF20aanxcpf show smooth, ongoing rises. On the other hand, ZTF20abrbeie, ZTF19aailpwl, ZTF20acvfraq and ZTF19aatubsj have MIR flares that peak within the timescale of the UV-optical light curve. MIR flares have been observed in TDEs \citep[e.g.][]{VanVelzen2016,jiang_infrared_2021}, BFFs and ANTs \citep[e.g.][]{Jiang2021,hinkle_mid-infrared_2024,Wiseman2023a} and are assumed to represent dust echoes from torus-like structures surrounding the black hole. The time lags between the UV-optical and MIR flares indicate the physical scale of the dust, the black body colour reveals its temperature, and the ratio of optical-to-MIR luminosity represents the dust covering fraction. 

We present the observed lags between the $g$-band and the $W1$, $W2$ light curves in Table \ref{tab:measured}. The lags are estimated at observer frame wavelengths but are corrected for time dilation. The roughly 6-month cadence of NEOWISE means that these lags are estimates, with uncertainty of $\pm 90$\,d. We present two lag measurements for each WISE band: the ``peak" lag, which is the time from $g$-band peak to the WISE peak; and the ``rise" lag, which we measure from the first rising optical detection to the first rising WISE detection. To estimate the time of peak brightness in each band we fit the light curves with Gaussian process (GP) regression \citep{Rasmussen_gaussian_2005} implemented via the \textsc{George} package \citep{Ambikasaran_fast_2015}. We use the squared exponential kernel with a scale length optimized via gradient-based methods, as per \citet{Pursiainen2020}. We constrain the MIR scale length to be greater than or equal to the optical, to account for the lower cadence of the MIR observations. In most cases we find the $W1$ and $W2$ peaks to be consistent with one another (i.e. within the 90\,d uncertainty). As described above, some events are yet to peak in the MIR such that their lags are lower limits. 

To estimate the rise lags, we simply take the time between the first rising point in the optical and MIR. For the optical we make use of ZTF forced photometry which often reveals rising light curves from before the first formal difference image detection. For the MIR, for which we do not have difference imaging, we estimate the host/AGN level by taking a weighted mean of the brightness before the first optical detection. None of the events show pre-flare variability in the MIR that is greater than the measurement uncertainties. The optical-MIR rise lag is then determined by the time of the first MIR point that lies significantly (i.e. (detection $-$ host) > host uncertainty). In many cases, this lag is clearly an upper limit set by the WISE cadence.
\begin{figure*}
    \centering
    \includegraphics[width=\textwidth]{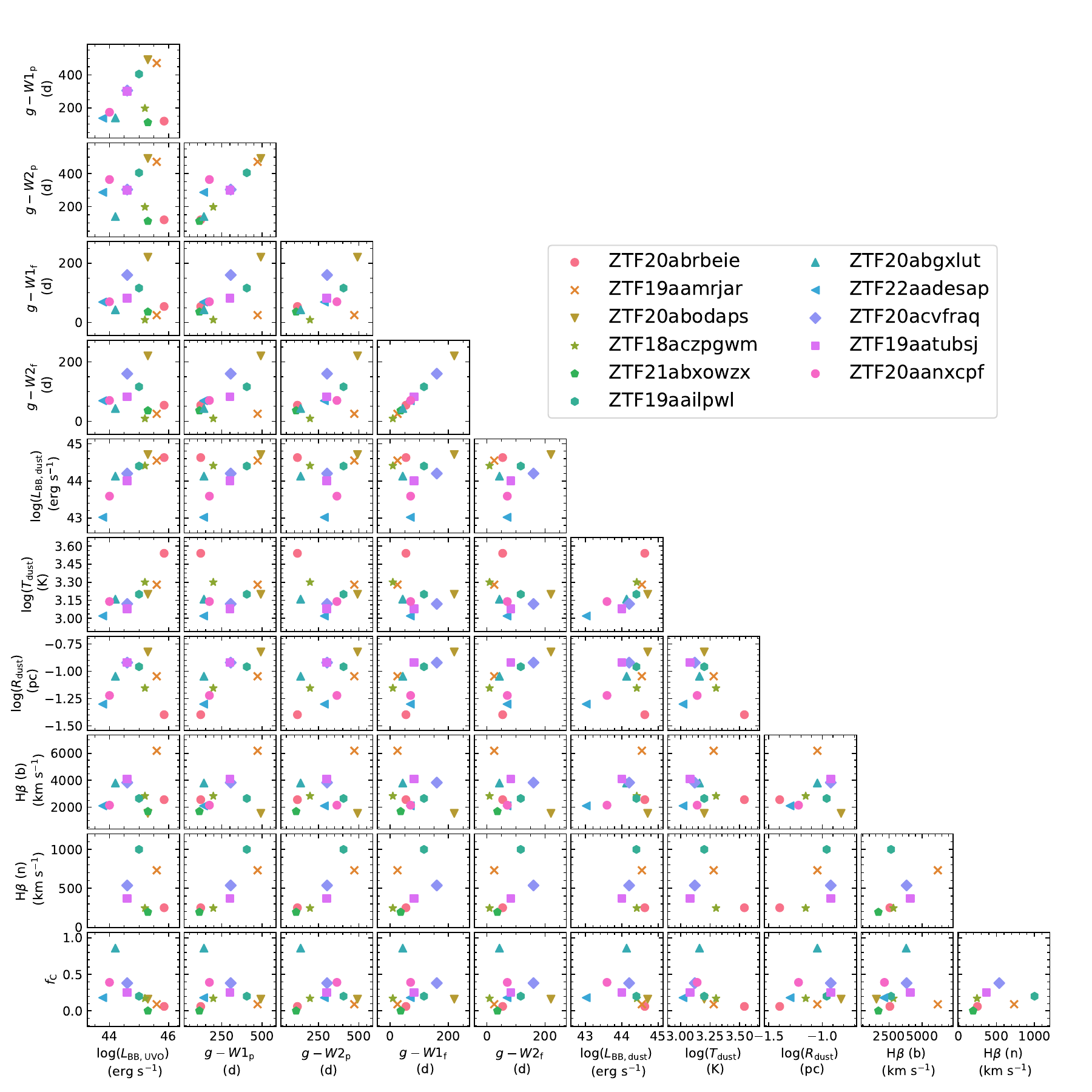}
    \caption{Correlations between optical flares, optical spectra, and dust light curves of the 11ANTs in the analysis sample.}
    \label{fig:dust_corr}
\end{figure*}

\subsubsection{Dust properties}
\label{subsubsec:dust_props}
We estimate the dust temperature, radius and luminosity by fitting the host-subtracted $W1$ and $W2$ data with a black body SED. We infer a dust covering fraction as the ratio between the estimated dust luminosity and the UV-optical luminosity. Due to having only two WISE filters and no near IR, the blackbody temperatures are poorly constrained -- formally the uncertainties are 100 per cent. Results are presented in Table \ref{tab:measured}. %To gain an estimate of the true uncertainties we {\bf will resample from the WISE photometry and refit the BB}. 
Our estimated dust properties are consistent with those measured for a sample of ANT MIR flares \citep{hinkle_mid-infrared_2024}. In particular, we compare the three events common to both samples (ZTF19aatubsj, ZTF20cvfraq, ZTF20aanxcpf) for which our temperature and effective radius measurements agree. The covering fractions are of similar magnitude to those reported by \citet{hinkle_mid-infrared_2024}. We note that dust emission is not a perfect black body: the main effects of allowing for an absorption coefficient would be a shift to lower inferred dust temperatures and larger distances from the source to the dust \citep{jiang_infrared_2021}, although neither of these affect our conclusions.

The effective dust temperatures lie within $1200\,{\rm K} < T_{\rm dust,peak} <2000\,{\rm K}$ which is consistent with the sublimation temperatures of graphite (1800\,K) and silicate (1500\,K) grains \citep{mor_dusty_2009,mor_hot_2012}. Correlations between the various dust parameters, lags, as well as the H$\beta$ widths and UV-optical black body luminosity are shown in Fig. \ref{fig:dust_corr}. We also measure the black body temperatures and radii from averaged pre-flare MIR fluxes. Assuming this pre-flare emission is dominated by circumnuclear dust illuminated by a bright UV-optical source (i.e. accretion disk), we can estimate the luminosity of this source, which we do following \citet{Petrushevska2023}. For the 7 sources with pre-flare MIR flux, we verify that the inferred pre-flare luminosity is smaller than the peak UV-optical luminosity by $\sim 1$ order of magnitude in each case.

We perform a simple Kendall $\tau$ test for correlation for all pairs of parameters. To determine which correlations are significant, we check for those with a $p$-value of 0.05 adapted for the look elsewhere effect for 50 parameter pairs, resulting in a $p$-value threshold of 0.001. Three pairs pass this threshold: the `first' lags and peak lags, although we note that these two lag measurements are not correlated with each other. This is likely due to the 6-month NEOWISE cadence. We also recover a correlation between dust luminosity the UV-optical luminosity ($\tau = 0.85$ $p = 0.00064$), further evidence that the MIR emission is indeed a dust echo of the UV-optical flare.

We also note that the $W1$ and $W2$ peak lags appear tentatively correlated with the width of narrow H$\beta$ ($\tau = 0.81$ $p = 0.01071$, with the high $p$-value due to some events not having a clear narrow feature) but not with that of broad H$\beta$. This may be evidence for the case where the narrow Balmer emission is caused by the reprocessing of kinetic energy of an outflow via the shock heating of a circumnuclear dust cloud, while the broad lines are from the outflow or accretion disk and thus do not correlate with the scale of the surrounding medium.

\subsection{Summary of properties}
In this section we have presented 11 examples of ANTs with $1$\,yr light curves. Seven are photometrically selected, and three spectroscopically. They are luminous with $L_{\rm BB}\geq10^{44}$\,erg\,s$^{-1}$; the six ANTs from the photometric selection that have constraints on their black body SEDs have $L_{\rm BB}\geq10^{45.1}$\,erg\,s$^{-1}$. All ANTs have strong, high equivalent-width Balmer lines. These lines are narrow compared to TDEs. The mean FWHM of the broad component of H$\beta$ in the ANTs is $\sim 2900$\,km\,s$^{-1}$ whereas for TDEs this is $>10^4$\,km\,s$^{-1}$ \citep{Charalampopoulos2022}.
All but one ANT has a strong MIR flare with a peak luminosity at least 10 per cent of the UV-optical black body luminosity, with temperature $1000 - 2000$\,K, with radius $\sim 0.1$\,pc and with a peak-to-peak lag from the optical $>100$\,d. All of these properties are consistent with a light echo from warm dust on the scales typically seen in AGN.

\section{Comparisons between the variability of ANTs and AGN}
\label{sec:SF}

\begin{figure}
    \centering
    \includegraphics[width=.5\textwidth]{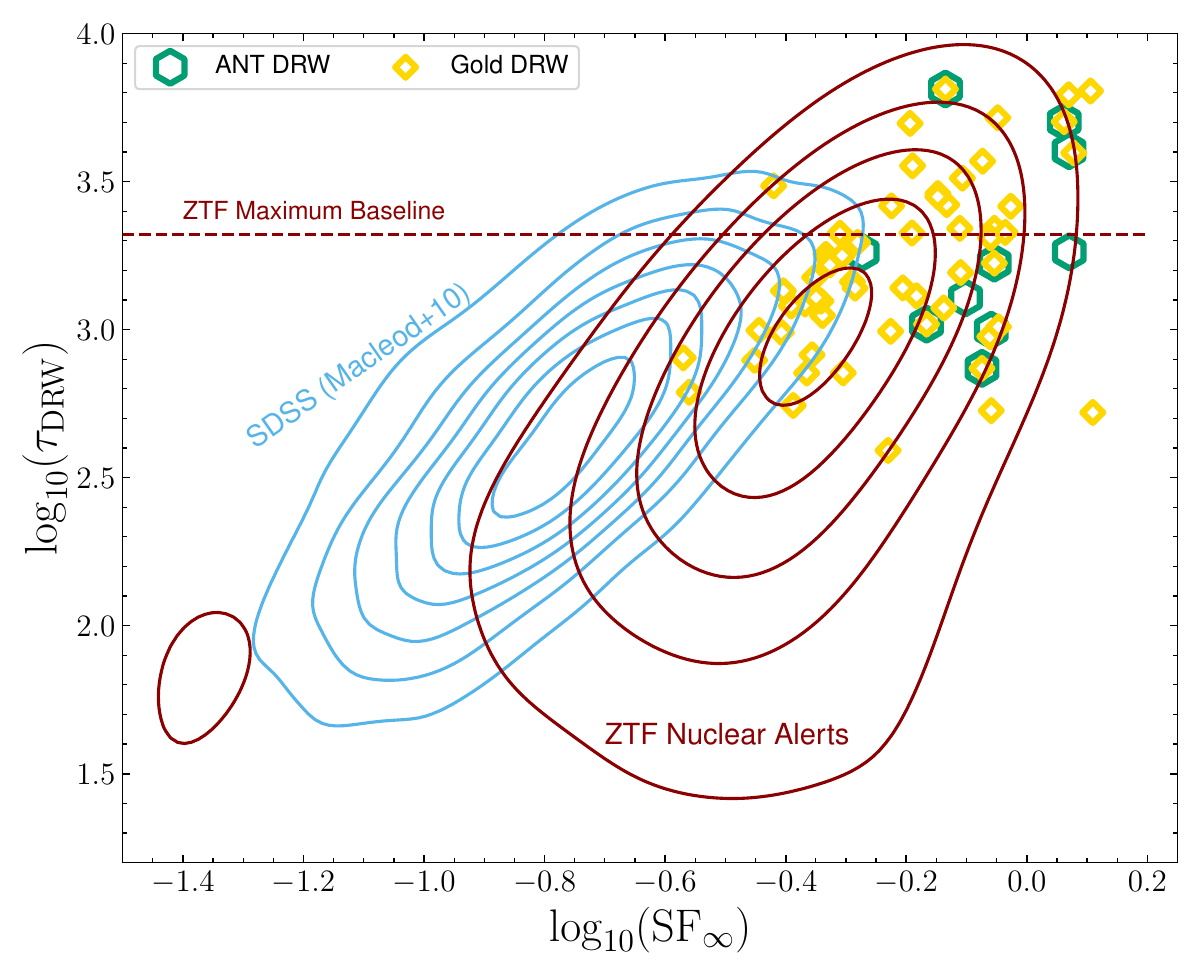}
    \label{fig:SF}
    \caption{Structure function of AGN compared to the variability shown by light curves selected to be smooth and high-amplitude. Structure functions for SDSS AGN and all $\sim 50,000$ ZTF nuclear alerts are shown in contours. Structure functions assuming a damped random walk (DRW) for our analysis sample of 11 ANTs, plus the gold sample of 59, are shown in open markers. }
    \includegraphics[width=.5\textwidth]{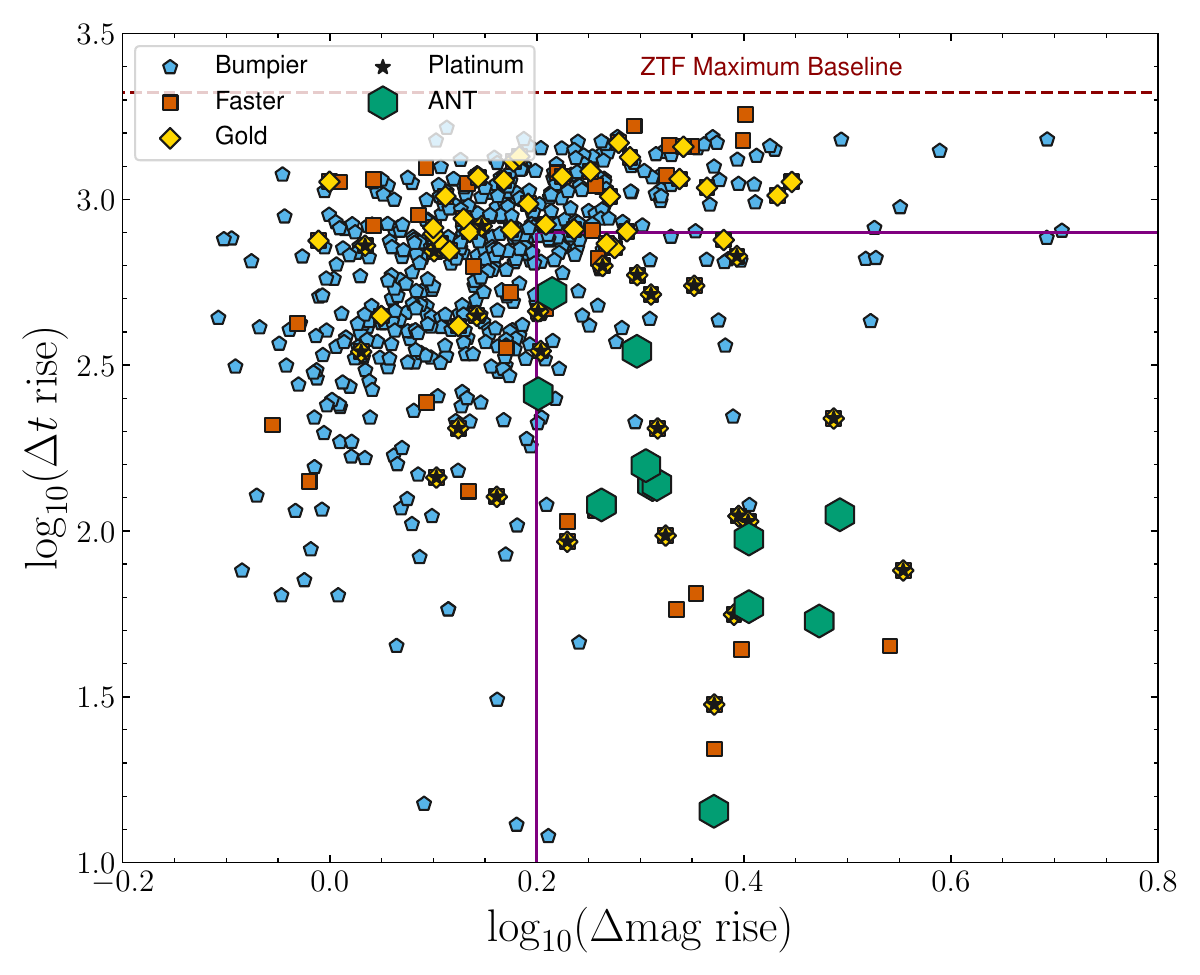}
    \label{fig:rise_amp}
    \caption{Rise times and rise amplitudes of all our sample variations. The purple box represents a selection of ANTs with 50\, per cent purity against AGN-like light curves. Note the x-axis is shifted to the right compared to Fig. \ref{fig:SF}}
    
\end{figure}

Distinguishing ANTs from the stochastic variability of AGN in a magnitude-limited photometric survey is non-trivial. An AGN that stays below the magnitude limit for several years and then fluctuates above that limit will mimic an ANT, which we assume is the case for 52/59 of the `gold' candidates that do not show smooth light curve evolution, as well as for the rest of the $\sim50,000$ passing the initial cuts. The separation of ANTs from AGN in real time is challenging, and will become more so with the increased depth of future surveys such as the Vera C. Rubin Observatory's Legacy Survey of Space and Time (LSST; \citealt{ivezic_lsst_2019}). Here we explore measures of light curve variability in order to learn how ANTs compare to the variability of typical AGN.

Variability of AGN is commonly described by a model-free {\it structure function} \citep[SF][]{Hughes_university_1992,Bauer_quasar_2009,Kozlowski_revisiting_2016}. The true SF, if observational noise has been accounted for, simply describes the root-mean-square variability about the mean AGN magnitude for a given timescale $\Delta t = t_i - t_j$. The observed SF is:
\begin{equation}
    {\rm SF}(\Delta t) = \sqrt{2\sigma^2_{\rm s} + 2\sigma^2_{\rm n} - 2{\rm cov}(s_i,s_j)}\,,
\end{equation}
where $\sigma^2_{\rm s}$ is the variance of the observed signal, $\sigma^2_{\rm n}$ is the noise, and ${\rm cov}(s_i,s_j)$ is the covariance between all data $s_i$, $s_j$ for that timescale $\Delta t$.

The SF can be parametrised by two power-laws, with index $\gamma>0$ for variability on timescales $\Delta t$ shorter than some break time $\tau$, and $\gamma=0$ at $\Delta t \to\infty$, i.e. that the amplitude of variability scales with timescale until $\tau$ where it becomes uncorrelated. The case of $\gamma=2$ corresponds to the damped random walk (DRW): in this case the SF can be described by the timescale $\tau_{\rm DRW}$ and the amplitude of the SF as $\Delta t \to\infty$, denoted ${\rm SF}_{\infty}$. The general population of AGN show a positive correlation between $\tau_{\rm DRW}$ and ${\rm SF}_{\infty}$ \citep[e.g.][]{macleod_modeling_2010}, such that objects with longer timescales tend to display higher-amplitude long-term variability. 

We assess the variability of the 11 ANTs in the analysis sample, the 59 `gold' candidates returned by our linear-decline, high-amplitude selection as well as the `faster' and `bumpier' samples. We also compute the SF for all ZTF nuclear transient candidates detected in the first year of ZTF, most of which are likely AGN. We restrict to these candidates so that we have at least a 5\,yr baseline. Instead of the difference-imaging magnitudes, which amplify differences with respect to some baseline, we obtain photometry from the ZTF Public Data Release 22\footnote{\url{https://www.ztf.caltech.edu/ztf-public-releases.html}} \citep{Masci2018} that includes any baseline AGN luminosity as well as the host galaxy contribution and can be compared directly to the well-studied SDSS sample. We fit the observer-frame $r$-band light curves with a DRW using the {\sc taufit}\footnote{\url{https://github.com/burke86/taufit}} package, fitting for the $\tau_{\rm DRW}$ and ${\rm SF}_{\infty}$ parameters. In an ideal case we would use rest frame light curves, but the majority of ZTF candidates do not have measured redshifts to allow for a $K$-correction. The results are shown in Fig. \ref{fig:SF}: SDSS AGN \citep{macleod_modeling_2010} and ZTF nuclear candidates are shown in blue and dark red contours, respectively while the SFs for ANTs in our gold sample are shown in open gold diamonds, and the SFs for ANTs are shown in open green hexagons. In general, DRW fits are unreliable if the observed baseline is not significantly longer than $\tau_{\rm DRW}$. For the 7229 ZTF nuclear light curves, the maximum possible baseline is $\sim1800$\,d, indicated by a horizontal line in the figure. Results above this line are unreliable while results below the line could be lower limits since longer baselines were not measureable. The ZTF candidates are all shifted to higher amplitude ${\rm SF}_{\infty}$ compared to the SDSS AGN, likely because all of these events are transient candidates detected in difference imaging and thus must be variable above the ZTF detection limit in the difference image (the $5\,\sigma$ threshold at $m_r\sim20$ is about 0.6\,mag.

\subsection{Rise time versus amplitude}
The 59 `gold' ANT candidates are selected to be smoothly evolving. Although the DRW is still a valid description of a low-amplitude, long timescale variability it is far from an ideal model for such transient variability. Nevertheless, the resulting ${\rm SF}_{\infty}$ all lie to the upper edge of the contour from ZTF, and are (by selection) larger than almost all \citet{macleod_modeling_2010} AGN. The analysis ANT sample is clearly shifted compared to the full `gold' sample which includes objects with shorter $\tau$ and smaller  ${\rm SF}_{\infty}$. To assess the significance of the largest flare, we measure the observer-frame rise time from first detection to peak brightness of the $r$-band light curve (not $K$-corrected) and plot these against the change in difference magnitude over the same timescale. These are shown as filled points in Fig. \ref{fig:rise_amp}, for the analysis sample of 11 plus the `gold', `bumpier' and `faster' selections. These measurements clearly indicate higher amplitude variability than measured by the DRW, and indicate a shorter timescale for that variability. There is a clear preference for the ANTs to have shorter rises from first detection to peak brightness compared to the `gold', `bumpier' and `faster' samples. This is despite the `faster' sample being selected for having a shorter decline, and indicates that most events picked up by that selection have slow rises and fast declines, indicative of stochastic variability rather than transient behaviour. 

We construct a region in rise-time vs rise amplitude space that includes all 11 ANTs from the analysis sample. This region is bounded by $\log_{10} (\Delta \mathrm{mag}~\mathrm{rise})>0.2$ (equivalent to 1.6\,mag) and $\log_{10} (\Delta t~\mathrm{rise})<2.9$ (equivalent to $\sim 800$\,d). We suggest that any light curves falling outside this range are likely due to stochastic AGN variability. Within this region are 9 `platinum' events that are not in our analysis sample. Of these, one is a peculiar TDE, two are SNe in a star-forming knots, and three are ANTs that pass our visual inspection but do not have spectra. Thus there are only 4 true AGN `contaminants' when applying these selection criteria and we advocate their use in future surveys.

\section{Discussion}
\label{sec:discussion}
\begin{table}
	\centering
	\caption{Black hole and stellar mass properties for the MOSFiT TDE model fit to 11 ANTs}
	\label{tab:mosfit}
	\begin{tabular}{llll} % four columns, alignment for each
            \hline
            ZTF ID & IAU Name & $\log_{10}(M_{\rm BH})$ & $M_*$\\
             & &${\rm M}_{\odot}$ & ${\rm M}_{\odot}$ \\
             \hline
            ZTF20abrbeie&  AT2021lwx & $8.32^{+0.01}_{-0.01}$ & $14.8^{+0.1}_{-0.3}$ \\\vspace{0.5mm}
            ZTF19aamrjar&  -         & $8.69^{+0.01}_{-0.01}$ & $90^{+7}_{-5}$ \\\vspace{0.5mm}
            ZTF20abodaps&  AT2020afep& - & - \\\vspace{0.5mm}
            ZTF18aczpgwm& AT2019kn & $7.5^{+0.05}_{-0.04}$ & $9.8^{+3.7}_{-6.1}$\\\vspace{0.5mm}
            ZTF21abxowzx&  AT2021yzu &$7.4^{+0.5}_{-0.7}$ & $2.5^{+1.1}_{-1.7}$\\\vspace{0.5mm}
            %ZTF20aaqtncr&  AT2021fez &$7.2^{+0.2}_{-0.1}$ & $8.5^{+4.8}_{-3.5}$\\\vspace{0.5mm}
            ZTF19aailpwl&  AT2019brs &$7.1^{+0.6}_{-0.4}$ & $1.5^{+0.3}_{-0.2}$\\\vspace{0.5mm}
            ZTF20abgxlut&  AT202oio &$7.04^{+0.15}_{-0.21}$&$0.28^{+0.22}_{-0.15}$\\ \vspace{0.5mm}
            ZTF22aadesap& AT2022fpx & $7.46^{+0.12}_{-0.05}$ & \textbf{$1.38^{+0.64}_{-0.86}$}\\\vspace{0.5mm}
            ZTF20acvfraq&  AT2020adpi&$7.0\pm0.1$ & $1.9^{+2.3}_{-0.8}$\\\vspace{0.5mm}
            ZTF19aatubsj&  AT2019fdr &$7.3\pm0.2$ & $4.5^{+5.2}_{-2.7}$\\\vspace{0.5mm}
            ZTF20aanxcpf&  AT2021loi &$6.5^{+0.3}_{-0.1}$ & $0.12\pm0.07$\\

\hline
            
  \end{tabular}
  
  \footnotesize{}
\end{table}
We have searched for analogues of the extremely luminous transient AT2021lwx, and have systematically identified a further sample of slow and smoothly evolving nuclear transients in the ZTF data stream. Of these transients, only one (ZTF19aailpwl/AT2019brs) has been published as a transient of interest. Another, ZTF21abxowzx/AT2021yzu, was publicly classified as an AGN. With this pipeline we identify four further events with photometric and spectroscopic properties consistent with ANTs. ZTF18aczpgwm/AT2019kn has broad, high equivalent-width emission lines as well as very strong iron complex. ZTF19aamrjar is the longest-lived of the sample and has the highest integrated radiated energy. Its spectrum resembles a quasar, and it has red-winged asymmetric Balmer lines, yet its photometric evolution is entirely smooth for over 2 years in the rest frame. ZTF20aaqtncr/AT2021fez is a lower amplitude flare whose emission lines resemble a type II AGN. ZTF20abodaps/AT2020afep rises on the timescale of a regular SN or TDE, but declines far, far slower. All of these transients display a MIR flare consistent with circumnuclear dust echoes. 
 
We supplemented these ANTs with three others observed by ZTF that do not have linear declines but whose spectroscopic properties do not fall into other astrophysical classes. Despite the discontinuous light curve shapes, the derived properties appear similar: the spectral shapes, Balmer line widths, UV-optical pseudo-bolometric luminosities and MIR flare characteristics are consistent with the linearly-declining sample.

%\subsection{Comparisons with known ANTs and TDEs}

\subsection{TDEs, CLAGNs, or something in between?}
The samples presented in this work add to a growing list of ANTs discovered in untargeted searches. Some previously published events (e.g. PS1-10adi, PS16dtm, Gaia16aaw,  ASASSN-17jz, Gaia18cdj) have similar smooth rises and declines to the ZTF sample. On the other hand, another set of events show differing light curves with either fast rises, bumpy declines, or plateaus (e.g. ASASSN-15lh, OGLE17aaj, AT2017bgt, ASASSN-18jd, Swift J221951-484240). Nevertheless, all but ASASSN-15lh are spectroscopically similar, dominated by strong narrow Balmer emission lines (and ASASSN-15lh has been claimed to be a TDE by \citealt{Leloudas2016}). These observations indicate that slow-moving gas is nearly ubiquitous in ANTs but that the geometry and/or central energy source may not be, leading to the different light curve shapes.

The physical nature of each individual ANT, and the class as a whole, is unclear. Many possible scenarios have been extensively discussed in the literature, and we have shown that any such scenario must readily explain strong narrow emission, large radii, and high dust covering fractions. Here we briefly address the most likely scenarios and how they relate to inferred properties of ANTs.

\subsubsection{Extreme instability in an existing accretion disk} 
Arguably the most simple explanation of ANTs is an intrinsic change in the structure of an existing accretion disk \citep[e.g.][]{sniegowska_possible_2020}. The existence of the disk before the flare is required in many ANTs which are found in NLSy1 galaxies or other AGN. This scenario is comprehensively addressed in \citet{Cannizzaro_extreme_2020} where it is shown that the rise timescales of ANTs (months) are far too short compared to the dynamical timescale of a classical accretion disk (>70 years), although see \citet{lipunova_fast_2024} for possible origins of flares due to thermal fluctuations in disks.

\subsubsection{Tidal disruption in an existing accretion disk}
One way of rapidly altering the accretion rate in an existing disk is via a TDE, a scenario proposed for many of the NLSy1 ANTs (e.g. AT2019fdr, \citealt{Frederick2021}; PS16dtm, \citealt{blanchard_ps16dtm_2017, Petrushevska2023}; PS1-10adi, \citealt{Kankare2017}) as well as some CLAGNs \citep[e.g.][]{Merloni_tidal_2015}. Predictions for the properties of such events are sparse \citep{Mckernan_starfall_2022,prasad_tidal_2024}. \citet{Chan_tidal_2019} show that the debris stream of a TDE interacts with the disk with a plethora of possible outcomes, dependent on the mass ratio of the debris and disk. They also indicate that a particularly heavy debris stream may hit the disk a second time, a possible explanation for the bumpier ANTs like ZTF19aatubsj/AT2019fdr and ZTF20aanxcpf.  

\subsubsection{Tidal disruption event of a massive star}
The ANTs presented here are extremely luminous and long-lived, properties indicative of a massive SMBH central engine. Black hole masses in canonical TDEs are found to be $10^5 - 10^7\,{\rm M}_{\odot}$ \citep{Mockler2019,Nicholl2022}, and solar mass stars can only be disrupted by black holes $\lesssim10^8\,{\rm M}_{\odot}$ \citep{Hills1975}. \citet{Wiseman2023a} and \citet{subrayan_scary_2023} showed that the light curve ZTF20abrbeie/AT2021lwx could be modelled by a $\sim 15\,{\rm M}_{\odot}$ star disrupted by a $8\times10^8\,{\rm M}_{\odot}$ black hole, and \citet{Wiseman2023a} claimed such an event was highly improbable due to the short lifetimes and low numbers of such massive stars. 

Here we fit the light curves of the analysis sample with the Modular Open Source Fitter for Transients \citep[MOSFiT][]{Guillochon2018} using the TDE model of \citet{Mockler2019}. We achieve reasonable fits to all the light curves except ZTF20abodaps/AT2020afep, for which the rise time is too fast compared to the decline time. The best fit stellar and black hole masses are presented in Table \ref{tab:mosfit}. The most eyecatching is ZTF19aamrjar, which requires $M_*\sim 90\,{\rm M}_{\odot}$ and $M_{\rm BH} = 5\times10^8\,{\rm M}_{\odot}$. Such a scenario is far beyond the scope of the MOSFiT models which derive from simulations of solar mass stars. All of the others have parameters more in line with what might be expected for extreme but plausible TDEs: $M_*< 3\,{\rm M}_{\odot}$ (with the exception of ZTF18aczpgwm/AT2019kn for which $M_*\sim 10\,{\rm M}_{\odot}$) and $M_{\rm BH}<10^8\,{\rm M}_{\odot}$. 

The mean black hole mass of our sample is $\log (M_{\rm BH}/{\rm M}_{\odot})=7.4$ with a standard deviation of 0.6\,dex. We compare these masses to those of 32 TDEs fit using the MOSFiT TDE model in \citet{Nicholl2022}, whose highest black hole mass is $\log (M_{\rm BH}/{\rm M}_{\odot})=7.2$ with a mean $\log (M_{\rm BH}/{\rm M}_{\odot})=6.6$. We perform a two-sample Kolmogorov-Smirnov (KS) test and conclude with a $p$-value of 0.0001 (i.e. 99.9\%, or $3.6\,\sigma$) that the ANT black hole masses are not drawn from the population of TDE black hole masses. We also compare to the 33 black hole masses of TDEs in \citet{Yao2023a} measured using the host galaxy bulge--black hole mass relation. Their mean mass is $\log (M_{\rm BH}/{\rm M}_{\odot})=6.5$ and although they extend to higher masses than the TDE black hole masses from MOSFiT, the KS test indicates the ANT population is different at $3\,\sigma$ significance.

Given that a significant number of the ANTs occurred in galaxies with likely non-negligible pre-existing accretion, we do not attempt to fit the host galaxy photometry to obtain a stellar mass to attempt to infer black hole masses.

\subsubsection{Tidal disruption of dense gas cloud}

The accretion of giant molecular clouds (GMCs) is a possible explanation to both ANTs as well as the rapid accretion rates necessary for the observed presence of very massive SMBHs at high redshift \citep[e.g.][]{Lin_rapid_2023}. The (partial) disruption and accretion of a GMC was suggested as an explanation for ZTF20abrbeie/AT2021lwx \citep{Wiseman2023a}. However, GMCs near the Milky Way Galactic centre have radii of several pc and masses $10^4-10^5\,{\rm M}_{\odot}$ \citep{Miyazaki_dense_2000} so that such events are predicted to have durations of $10^4 - 10^5$\,yr for a $10^6\,{\rm M}_{\odot}$ SMBH \citep{Alig_simulations_2011,Alig_numerical_2013}, and even longer for the more massive cases. Nevertheless, the near-ubiquitous MIR flares in our sample are strong evidence that dust and molecular gas is present at $\sim 0.1-1$\,pc from the black hole. On these scales the clouds are smaller \citep{Hsieh_circumnuclear_2021}, and around AGN it is known that the cold and warm dust and molecular gas constituting the torus is clumpy \citep{Krolik_molecular_1988} and turbulent \citep[e.g]{wada_molecular_2009, Hoenig_donuts_2013} with many cloud-cloud collisions \citep[e.g.][]{Beckert_dynamical_2004}. A scenario could thus be imagined where such a cloud-cloud collision results in a direct encounter between a small, dense cloud and the SMBH, resulting in super-Eddington accretion and an ANT-like flare. 

\subsection{Volumetric rate of ANTs}
The intrinsic rate of transient events provides constraints on the physical mechanisms and plausible scenarios causing them. For example, for ANTs to be caused by high stellar mass TDEs, the rates should agree with the expected population of high-mass stars in SMBH loss cones. Such rate measurements are important for the inclusion of ANTs in simulations of galaxies and their nuclei, the understanding of stellar dynamics and black hole mass distributions \citep[e.g.][]{Stone_rates_2016} as well as for explaining the chemical abundances of AGN \citep[e.g.][]{Kochanek_abundance_2016}. 
We estimate the volumetric rate of ANTs by using the seven events in the analysis sample that were selected by our pipeline. We limit the calculation to $z<1$, the highest redshift event (ZTF20abrbeie/AT2021lwx). For each event we calculate the maximum redshift that it would have passed our cuts, using the black body properties from Table \ref{tab:bb_measurements}. Using that maximum redshift, we then weight each transient by the maximum observable volume, $1/V_{\rm max}$ \citep{Schmidt1968}. We assume the sky area of the public survey to be $15,000$\,deg$^2$ \citep{ho_search_2023}, and the survey time to be 4.77 yr (we cut the selection at 31/12/2023/MJD 60309 so for the transient to last 1 yr, it must be first detected by MJD 59943). Counting the seven events in this manner, we estimate a lower limit on the rate $\gtrsim 5\times10^{-11}$\,Mpc$^{-1}$\,yr$^{-1}$. Removing ZTF20aaqtncr/AT2021fez, which we classified as Unclear, gives a rate of $\gtrsim 
 3\times10^{-11}$\,Mpc$^{-1}$\,yr$^{-1}$. We note that these measurements are $\sim50$ times greater than those for `extreme' ANTs (ENTs) measured by \citet{hinkle_extreme_2024}, a sample of three with which we share one common event (ZTF20abrbeie/AT2021lwx). These rates are consistent with a direct extrapolation of the measured TDE luminosity function (i.e. rate per unit peak luminosity) in ZTF measured by \citet{Yao2023a}. However, above the Hills mass of $M_{\rm BH}\geq10^8\,{\rm M}_{\odot}$ for a solar mass star the TDE rate should be heavily suppressed \citep[e.g.][]{Stone_rates_2020}, and indeed such a suppression is observed \citep{van_velzen_late-time_2019,Mockler2019,Nicholl2022}. \citet{Wiseman2023a} estimated that the rate of TDEs from $\sim15\,{\rm M}_{\odot}$ stars should be $10^{-6}$ that of solar mass stars based on the slope of the initial mass function (IMF) and the lifetime of massive stars. Applying this fraction to the observed volumetric TDE rate of $3.1\times10^{-7}$\,Mpc$^{-1}$\,yr$^{-1}$ \citep{Yao2023a} we might expect $\sim 10^{-13}$\,Mpc$^{-1}$\,yr$^{-1}$ from $\sim15\,{\rm M}_{\odot}$ stars. The TDE model fits to our ANT light curves returned stellar masses between $0.12 - 90\,{\rm M}_{\odot}$, with seven in the range $1.5 - 10\,{\rm M}_{\odot}$. Based upon the arguments of lifetimes and the IMF would expect significantly more TDEs from stars in this mass range than $\sim15\,{\rm M}_{\odot}$. The observed rate $\gtrsim 3\times10^{-11}$\,Mpc$^{-1}$\,yr$^{-1}$ is thus consistent with the expectation for TDEs of stars in the $1.5 - 10\,{\rm M}_{\odot}$ mass range. The increased depth of the Vera C. Rubin Observatory's Legacy Survey of Space and Time \citep[LSST][]{ivezic_lsst_2019} will enhance the rate of observed TDEs \citet{bucar_bricman_rubin_2023} while hundreds will be spectroscopically classified by the Time Domain Extragalactic Survey (TiDES; \citealt{Frohmaier2025}) on the 4-metre Multi-Object Spectrograph Telescope. These observations will allow a detailed measurement of the rate of massive-star TDEs and ANTs.

\subsection{Uniqueness of AT2021lwx}
Despite a thorough search, no events were found with observed properties that are comparable properties to AT2021lwx. Firstly, every event discovered using the linear fit was found with a clear host galaxy, with AT2021lwx itself the only hostless event passing the cuts. No other event in our sample has a luminosity as low as the limit of $M_r>-21$\,mag for the host of AT2021lwx. Assuming a black body SED for the transient, two events (ZTF19aamrjar and ZTF20abodaps/AT2020afep) from the analysis sample come within $\sim 0.3$\, dex of the peak UV-optical pesudo-bologmetric luminosity of AT2021lwx ($\log(L_{\rm{bol,max}}/{\rm erg\,s}^{-1})$=45.85). The light curve of ZTF19aamrjar is the most similar to AT2021lwx, and since it evolves much more slowly, the integrated energy release is larger than AT2021lwx. The spectrum is however vastly different: the Balmer line profiles are broad and asymmetric, and there is strong narrow oxygen. ZTF20abodaps/AT2020afep has a similar spectrum, with no oxygen present. However, the sharp rise time is inconsistent with AT2021lwx.

\section{Conclusions}
\label{sec:conclusion}
In this paper we have presented a systematically selected sample of ambiguous nuclear transients with smooth, slowly evolving optical light curves. The events appear to span the continuum between Bowen fluorescence flares and the extreme end of AGN variability. A summary of the key findings is below:
\begin{itemize}
    \item We find eight nuclear events with long-duration ($>300$\,d), smooth, luminous ($L_{\rm BB}>10^{44.7}$\,erg\,s$^{-1}$) light curves similar to ZTF20abrbeie/AT2021lwx including ZTF20abrbeie/AT2021lwx itself.
    \item We find three additional transients with bumpier light curves whose spectra do not resemble regular AGN or SNe.
    \item The light curve properties are heterogeneous, with rise times ranging from 20\,d to 1\,yr. 
    \item The spectra are also heterogeneous, ranging from quasar-like (ZTF19aamrjar), through NLSy1-like (ZTF18aczpgwm/AT2019kn) through to Balmer emission only (ZTF20abodaps/AT2020afep and ZTF20abrbeie/AT2021lwx).
    \item Mid-infrared flares are near ubiquitous, implying the presence of circumnuclear dust.
    \item Modelling the light curves as TDEs, the inferred black hole masses and stellar masses are beyond the ranges observed for canonical TDEs. The volumetric rate of $\gtrsim 3\times10^{-11}$\,Mpc$^{-1}$\,yr$^{-1}$ is consistent with rough predictions of disruptions of intermediate high-mass stars.
    \item ANTs have rise times longer than and amplitudes greater than regular AGN variability. We define a region of rise-time versus flare amplitude parameter space for a complete selection of ANTs that is $\sim50$ per cent contaminated by AGN.
    
\end{itemize}
If these ANTs, along with those compiled from other surveys, are indeed TDEs then the emission mechanism for the bulk of the radiation must be different to the canonical TDE of a $1\,{\rm M}_{\odot}$ star around a $10^5 - 10^7\,{\rm M}_{\odot}$ black hole in order to produce the markedly different optical spectra.

We expect to observe $>100$ AT2021lwx-like ANTs with LSST and will be able to statistically model their light curve, spectral, and host galaxy properties.

\section*{Acknowledgements}
We thank the referee for their thoughtful comments which have greatly improved the work. We thank Karri Koljonen for providing the spectra and reduced UVOT data for ZTF22aadesap/AT2022fpx.
We thank S. Smartt and the Pan-STARRS team for providing data points of ZTF19aamrjar, ZTF20abodaps, ZTF18aczpgwm, ZTF19aatubsj, ZTF20aanxcpf to support the figures in the appendix. We are grateful to Matt Stepney and Daniel Kynoch for their advice and insights on AGN spectra and Sandra Raimundo for discussions about CLAGNs.

PW acknowledges support from the Science and Technology Facilities Council (STFC) grant ST/R000506/1.
IA acknowledges support from the European Research Council (ERC) under the European Union’s Horizon 2020 research and innovation program (grant agreement number 852097), from the Israel Science Foundation (grant number 2752/19), from the United States - Israel Binational Science Foundation (BSF; grant number 2018166), and from the Pazy foundation (grant number 216312).
LG acknowledges financial support from the Spanish Ministerio de Ciencia e Innovaci\'on (MCIN), the Agencia Estatal de Investigaci\'on (AEI) 10.13039/501100011033, the European Social Fund (ESF) ‘Investing in your future’, the European Union Next Generation EU/PRTR funds under the PID2020-115253GA-I00 HOSTFLOWS project, the 2019 Ram\'on y Cajal programme RYC2019-027683I, the 2021 Juan de la Cierva programme FJC2021-047124-I, Centro Superior de Investigaciones Cient\'ificas (CSIC) under the PIE project 20215AT016, and the program Unidad de Excelencia Mar\'ia de Maeztu CEX2020-001058-M.
MN was supported by the European Research Council (ERC) under the European Union’s Horizon 2020 research and innovation programme (grant agreement no. 948381) and by UK Space Agency Grant No.~ST/Y000692/1. TP acknowledges the financial support from the Slovenian Research Agency (grant P1-0031). DI acknowledges funding provided by the University of Belgrade - Faculty of Mathematics (the contract \textnumero 451-03-66/2024-03/200104) through the grant of the Ministry of Science, Technological Development and Innovation of the Republic of Serbia. MJG acknowledges support from the National Science Foundation (grant AST-2108402).

Lasair is supported by the UKRI Science and Technology Facilities Council and is a collaboration between the University of Edinburgh (grant ST/N002512/1) and Queen’s University Belfast (grant ST/N002520/1) within the LSST:UK Science Consortium.

%%%%%%%%%%%%%%%%%%%%%%%%%%%%%%%%%%%%%%%%%%%%%%%%%%
\section*{Data Availability}

All of the ZTF lightcurves from Lasair are freely available, for example: 
 \url{https://lasair-ztf.lsst.ac.uk/object/ZTF19aamrjar}, and others following the same pattern.
 
All photometric and spectroscopic data presented in this manuscript is available at \url{https://github.com/wisemanp/ANTs-Nest}.



\bibliographystyle{mnras}
\bibliography{ZoteroBibOffline}
%\bibliography{zotero_offline} % if your bibtex file is called example.bib

%\begin{thebibliography}{99}
%\end{thebibliography}

%%%%%%%%%%%%%%%%%%%%%%%%%%%%%%%%%%%%%%%%%%%%%%%%%%

%%%%%%%%%%%%%%%%% APPENDICES %%%%%%%%%%%%%%%%%%%%%

\appendix

\section{Varying selection criteria}
\label{sec:app_selection}
Here we test our selection criteria with three variations on the nominal `gold' selection described in Section \ref{subsec:selection}.

\subsection{Decline time longer than rise time}
The first variation is to impose a requirement that the decline time is longer than the rise time, as expected for accretion-powered transients. 21 transients pass this cut, which we label the `platinum' selection.

\subsection{Faster light curves}
The second variation is to relax the light curve duration to $\geq 0.5$\,yr. We also relax the number of detections to 25. This allows for faster-evolving events, and events with limited visibility (e.g. a 150\,d rise was unobservable from Palomar but a subsequent 200\,d decline was observed). It also means the contamination from regular transients is increased. 95 transients pass this `faster' selection, of which 11 pass visual inspection having not already been included in the gold sample. Of the 11, there are three spectroscopically classified SNe~IIn: ZTF19abpvbzf appears in a knot of star formation in an irregular galaxy, ZTF22aadesjc appears visually off-centre in a dwarf galaxy, but ZTF22aaurfwa=SN2022hem is only classified by a low-resolution spectrum and may be an ANT although the post-peak colour evolution is more consistent with the SN~II classification. Four objects are spectroscopically classified TDEs: ZTF18abxftqm=AT2018hco \citep{van_velzen_classification_2018}, ZTF20abnorit=AT2020ysg \citep{hammerstein_final_2023}, ZTF20achpcvt =AT2020vwl \citep{hammerstein_ztf_2021}. ZTF23aadcbay=AT2023cvb \citep{Yao_ztf_2023}. ZTF18aaqkcso is a type Ia SN. In Fig \ref{fig:lcs_bumpy} we show the light curve of ZTF20aaurfwa/SN2022hem, along with ZTF20abnorit/AT2020ysg which is classified as TDE-featureless and thus is ambiguous. 

\subsection{Bumpier light curves}
\begin{figure*}
    \centering
    \includegraphics[width=\textwidth]{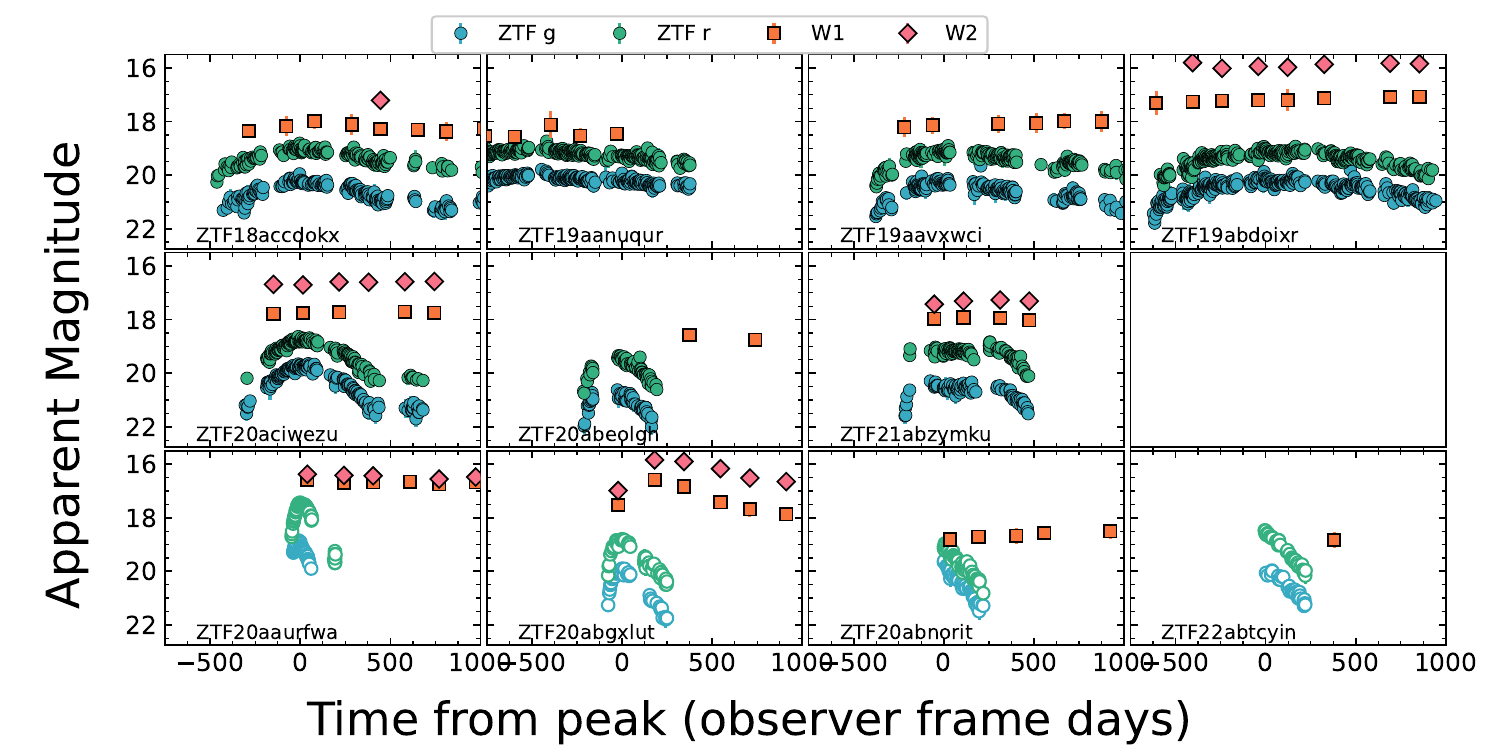}
    \caption{Optical and MIR light curves of the 7 ANTs passing `bumpier' cuts (filled circles) and the 4 passing `faster' cuts (open circles).}
    \label{fig:lcs_bumpy}
\end{figure*}

The third variation is to relax the constraint on the linear fit to $R^2\geq0.5$, allowing for noisier and/or bumpier light curves. 550 events pass these criteria, many of which are regular AGN-like variability. 10 of these events are labelled ANTs in visual inspection. Of these 10, three are known QSOs while one (ZTF20acklcyp) is spectroscopically classified as a SN~IIn (SN2020xkx). The remaining six are shown in Fig. \ref{fig:lcs_bumpy}.

\subsection{ANTs with no spectra}
\begin{figure*}
    \centering
    \includegraphics[width=.675\textwidth]{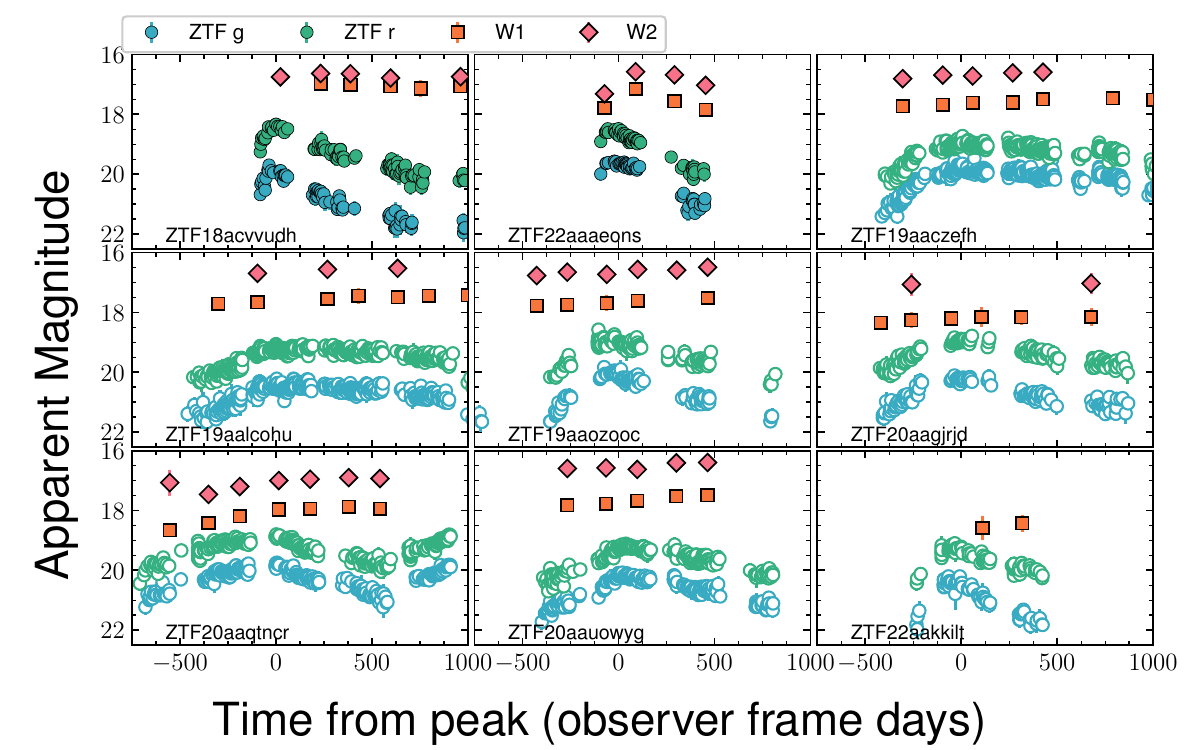}
    \caption{Optical and MIR light curves of the 2 ANTs passing `gold' cuts but without spectroscopic redshift (filled circles) and the 4 passing gold cuts but in known QSOs (open circles).}
    \label{fig:lcs_nospec_or_qso}
\end{figure*}

ZTF18acvvudh and ZTF22aaaeons passed our gold (and platinum) selection cuts and visual inspection, but do not have spectra or redshifts and thus are not included in the main analysis. Their light curves are shown in filled circles in Fig. \ref{fig:lcs_nospec_or_qso}.

\subsection{ANTs in known QSOs}
Four events, ZTD19aaczefh, ZTF19aalcohu, ZTF20aagjrjd and ZTF22aakkilt show ANT-like light curves but occurred in known spectroscopic quasi-stellar objects (QSOs, i.e. high-z AGNs). In contrast to our analysis sample, they show minimal MIR evolution.

%%%%%%%%%%%

\section{Full light curves}
\begin{figure} 
     \centering 
     \includegraphics[width=.48\textwidth]{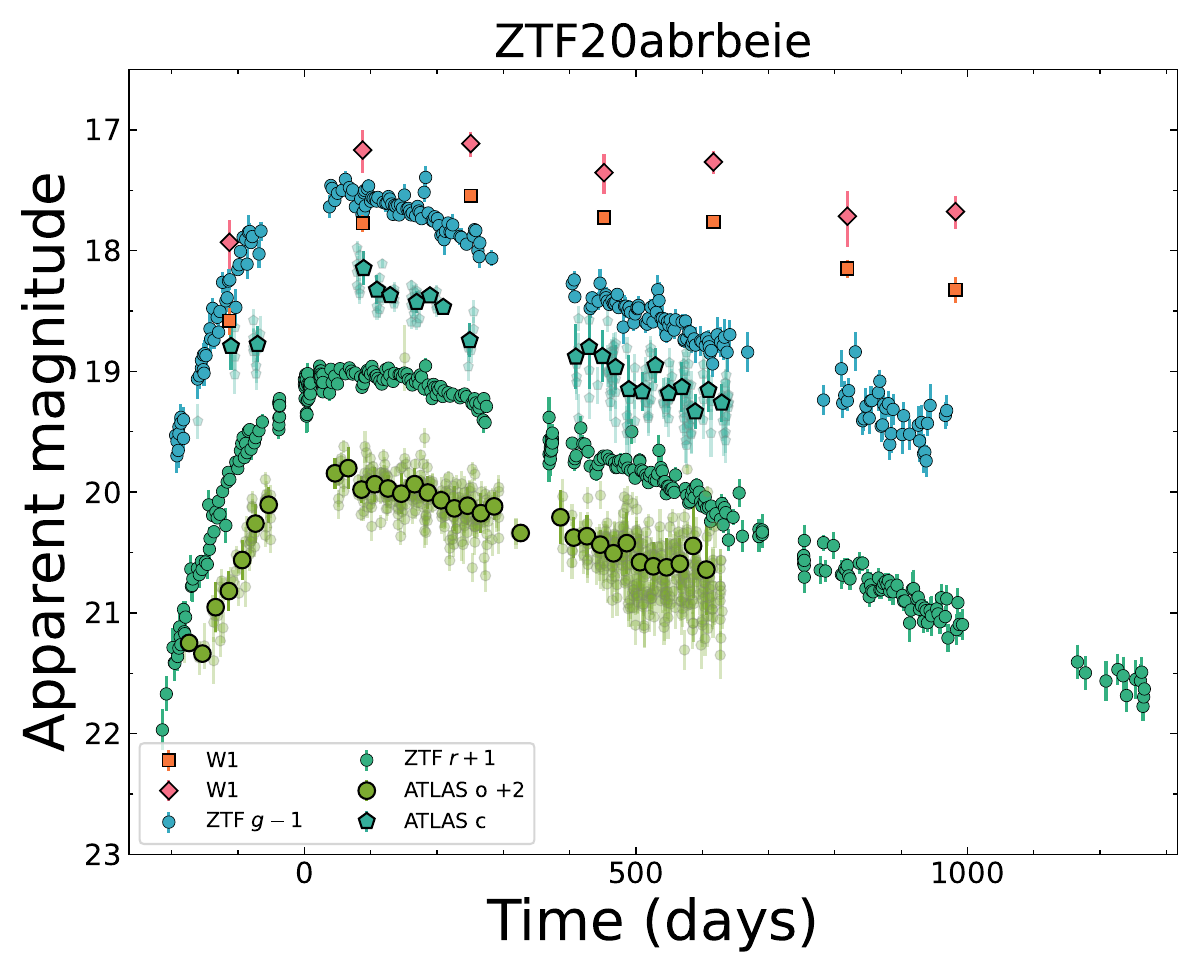} 
     \caption{Observer frame light curve of ZTF20abrbeie relative to $r$-band maximum light.} 
     \label{fig:ZTF20abrbeie_lc_allbands} 
 \end{figure}

\begin{figure} 
     \centering 
     \includegraphics[width=.48\textwidth]{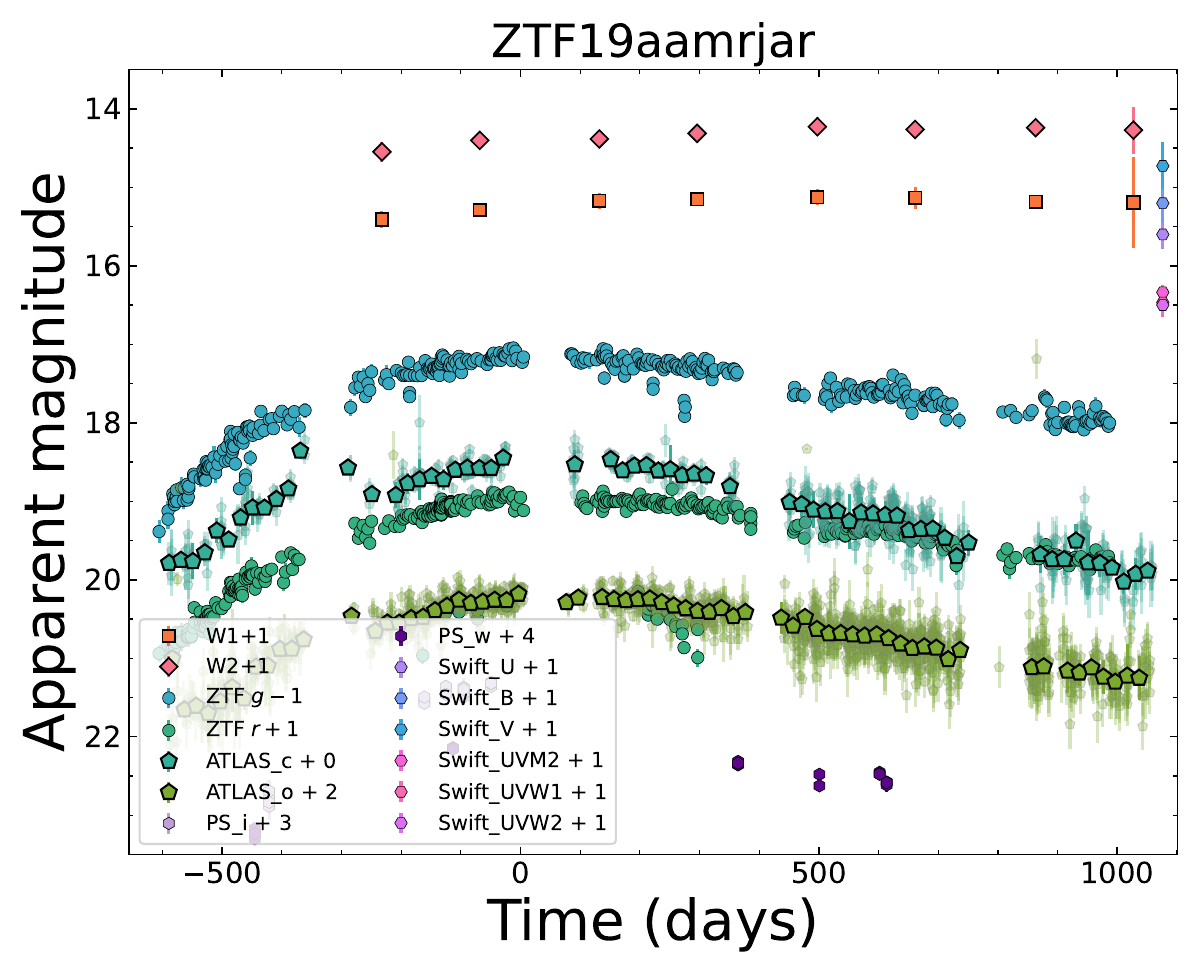} 
     \caption{Observer frame light curve of ZTF19aamrjar relative to $r$-band maximum light.} 
     \label{fig:ZTF19aamrjar_lc_allbands} 
 \end{figure}

\begin{figure} 
     \centering 
     \includegraphics[width=.48\textwidth]{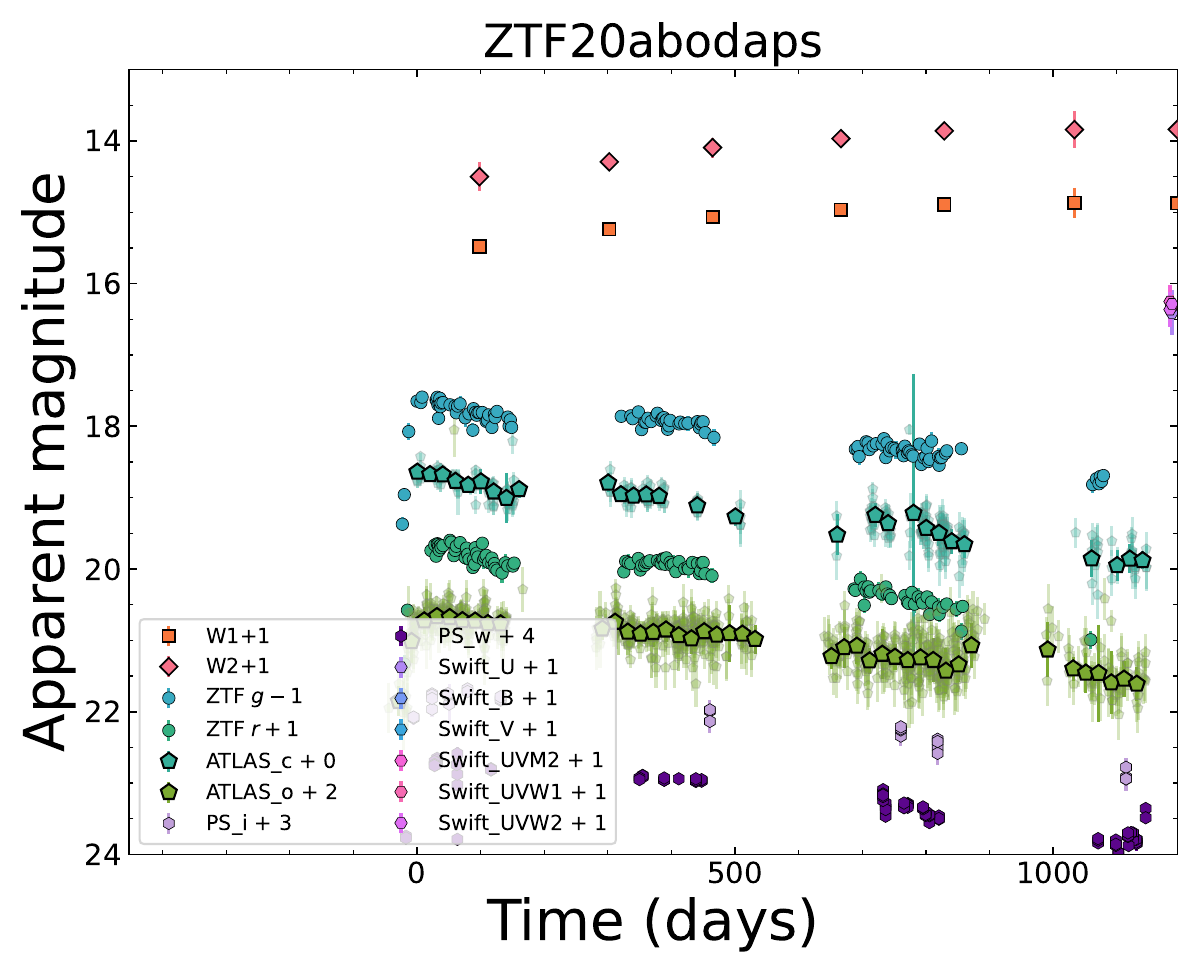} 
     \caption{Observer frame light curve of ZTF20abodaps relative to $r$-band maximum light.} 
     \label{fig:ZTF20abodaps_lc_allbands} 
 \end{figure}

\begin{figure} 
     \centering 
     \includegraphics[width=.48\textwidth]{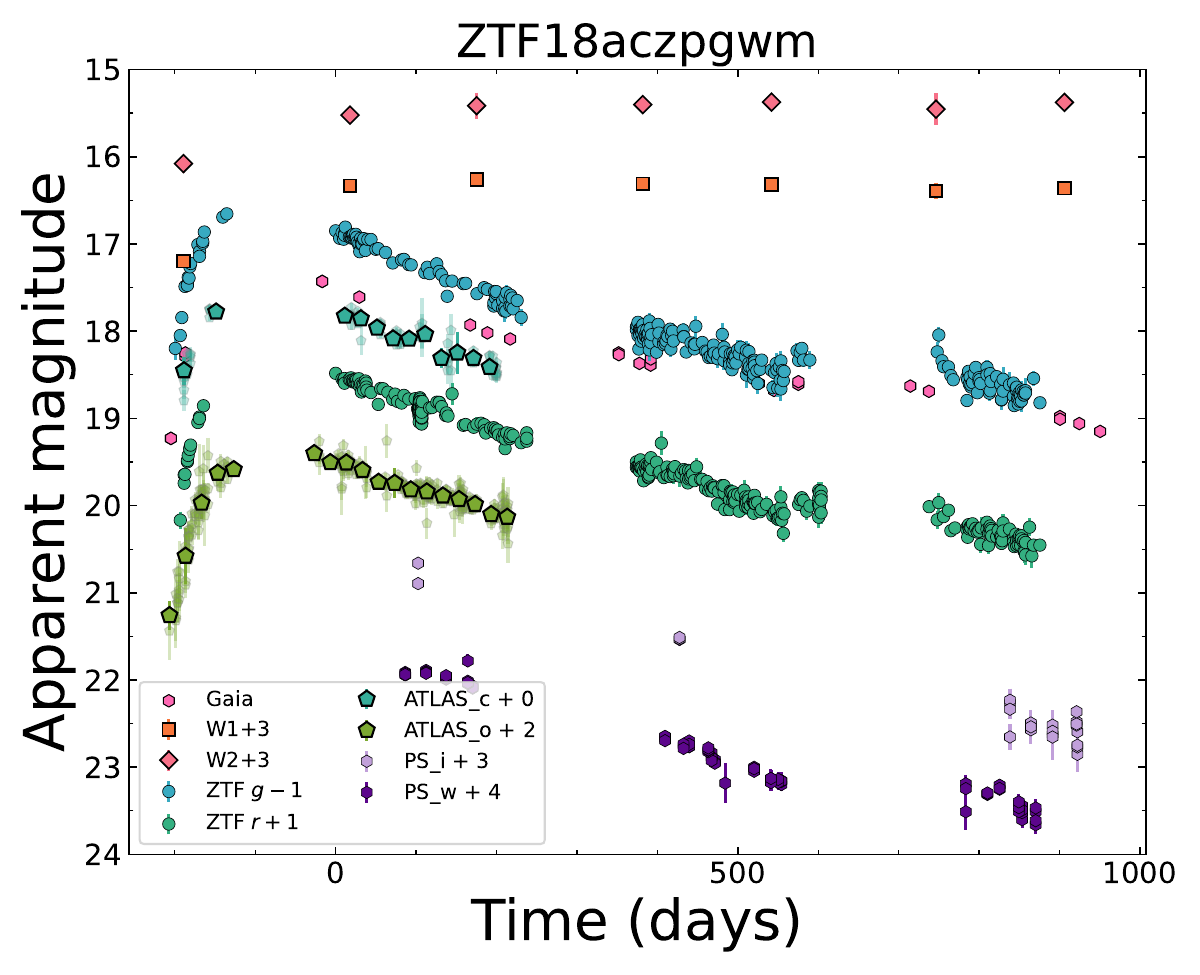} 
     \caption{Observer frame light curve of ZTF18aczpgwm relative to $r$-band maximum light.} 
     \label{fig:ZTF18aczpgwm_lc_allbands} 
 \end{figure}

\begin{figure} 
     \centering 
     \includegraphics[width=.48\textwidth]{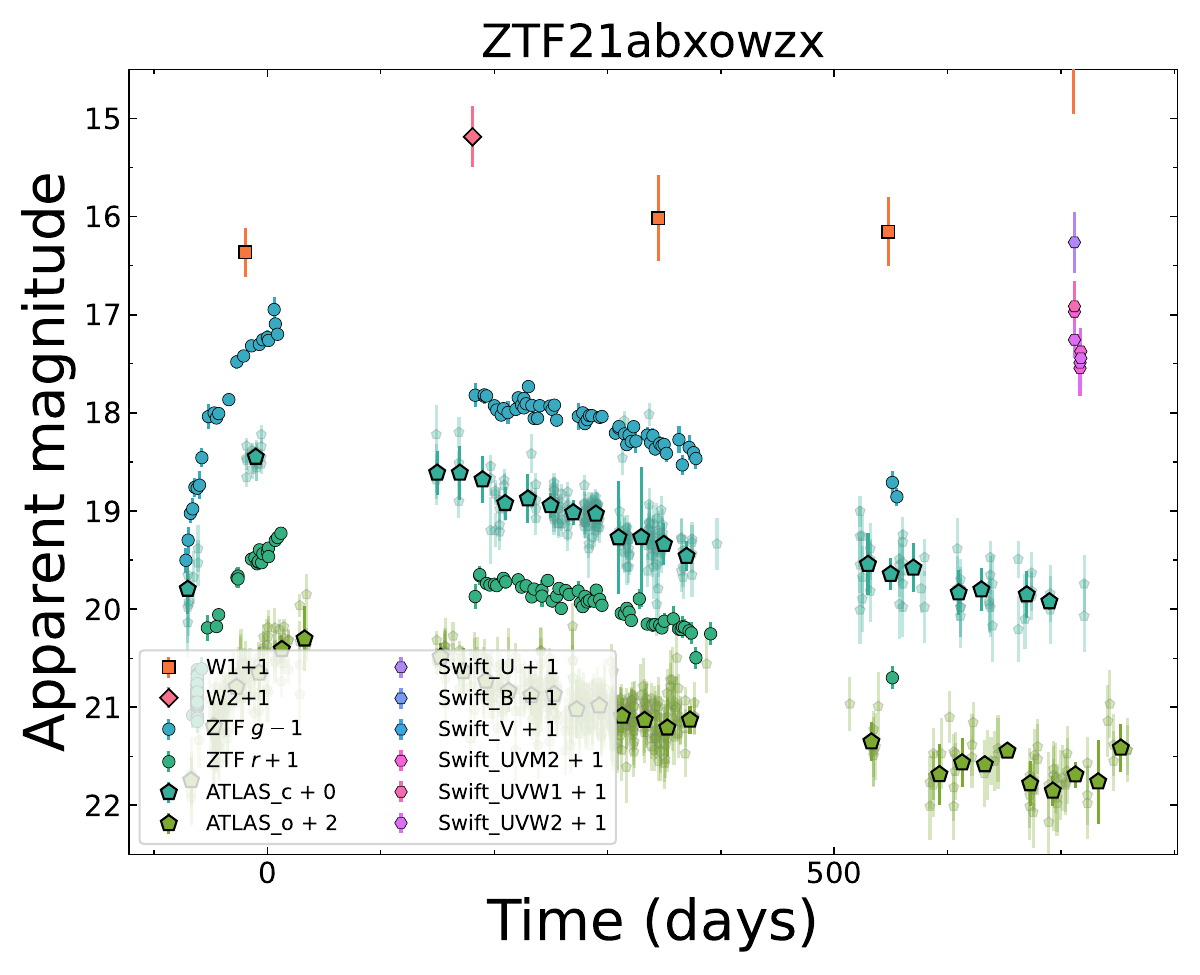} 
     \caption{Observer frame light curve of ZTF21abxowzx relative to $r$-band maximum light.} 
     \label{fig:ZTF21abxowzx_lc_allbands} 
 \end{figure}

\begin{figure} 
     \centering 
     \includegraphics[width=.48\textwidth]{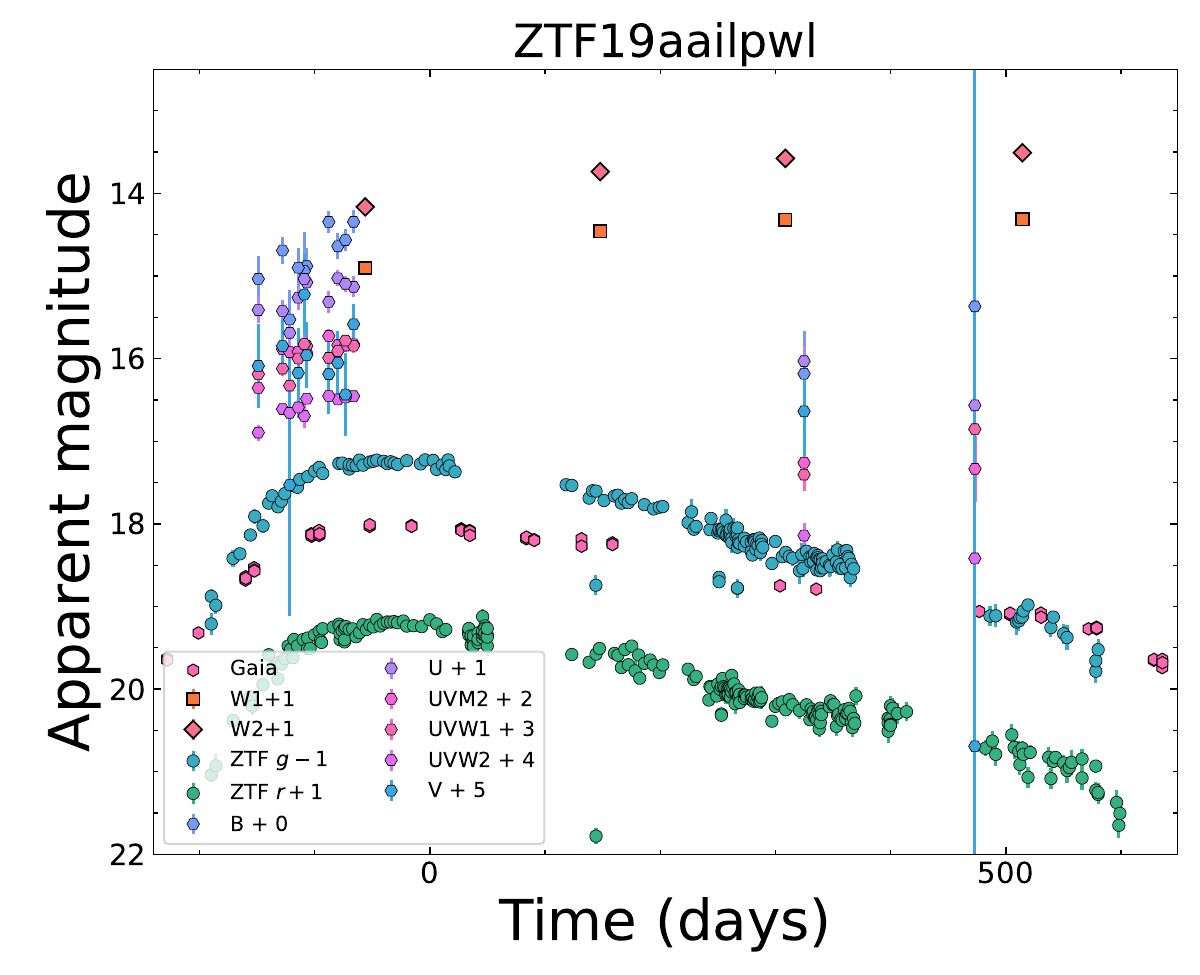} 
     \caption{Observer frame light curve of ZTF19aailpwl relative to $r$-band maximum light.} 
     \label{fig:ZTF19aailpwl_lc_allbands} 
 \end{figure}

\begin{figure} 
     \centering 
     \includegraphics[width=.48\textwidth]{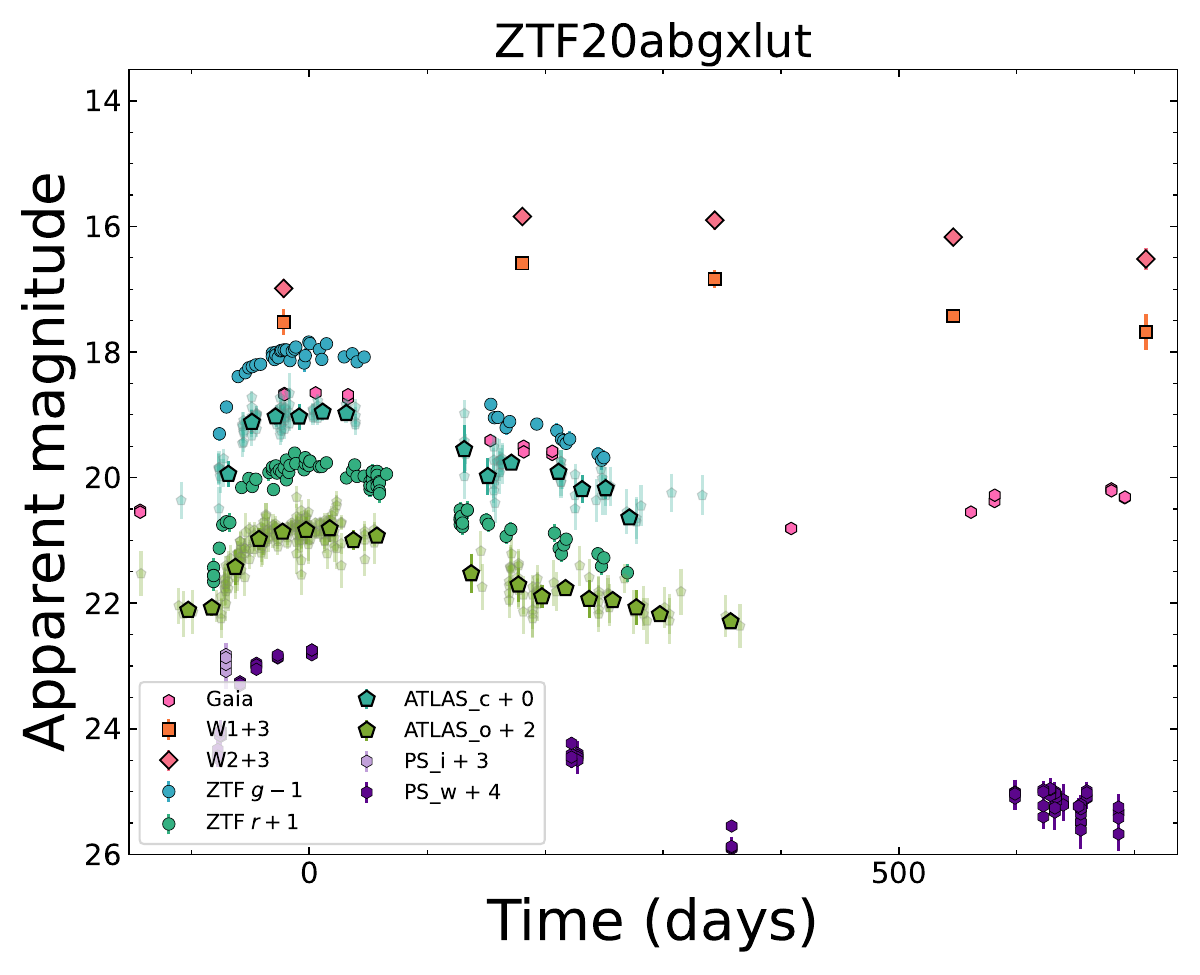} 
     \caption{Observer frame light curve of ZTF20abgxlut relative to $r$-band maximum light.} 
     \label{fig:ZTF20abgxlut_lc_allbands} 
 \end{figure}

\begin{figure} 
     \centering 
     \includegraphics[width=.48\textwidth]{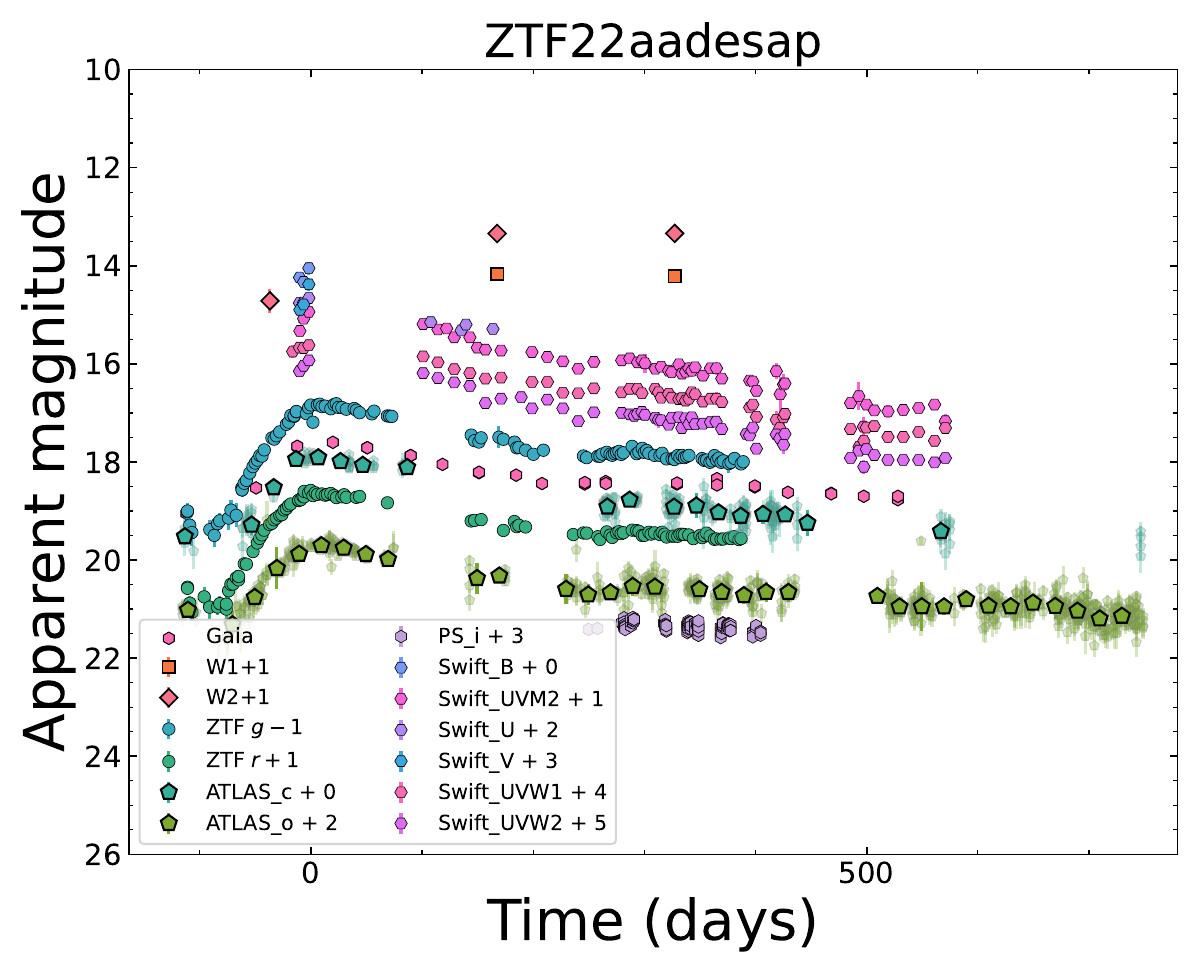} 
     \caption{Observer frame light curve of ZTF22aadesap relative to $r$-band maximum light.} 
     \label{fig:ZTF22aadesap_lc_allbands} 
 \end{figure}

\begin{figure} 
     \centering 
     \includegraphics[width=.48\textwidth]{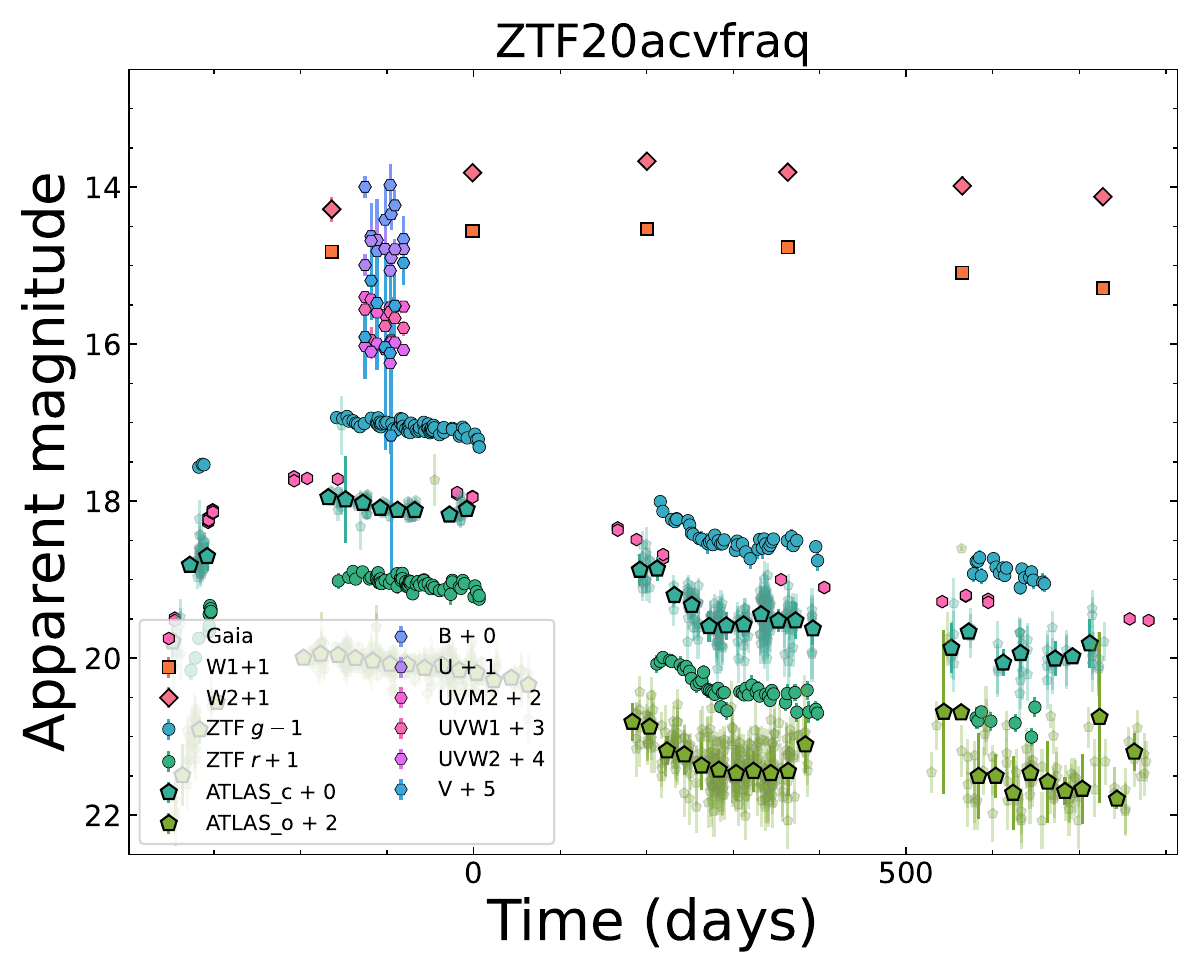} 
     \caption{Observer frame light curve of ZTF20acvfraq relative to $r$-band maximum light.} 
     \label{fig:ZTF20acvfraq_lc_allbands} 
 \end{figure}

\begin{figure} 
     \centering 
     \includegraphics[width=.48\textwidth]{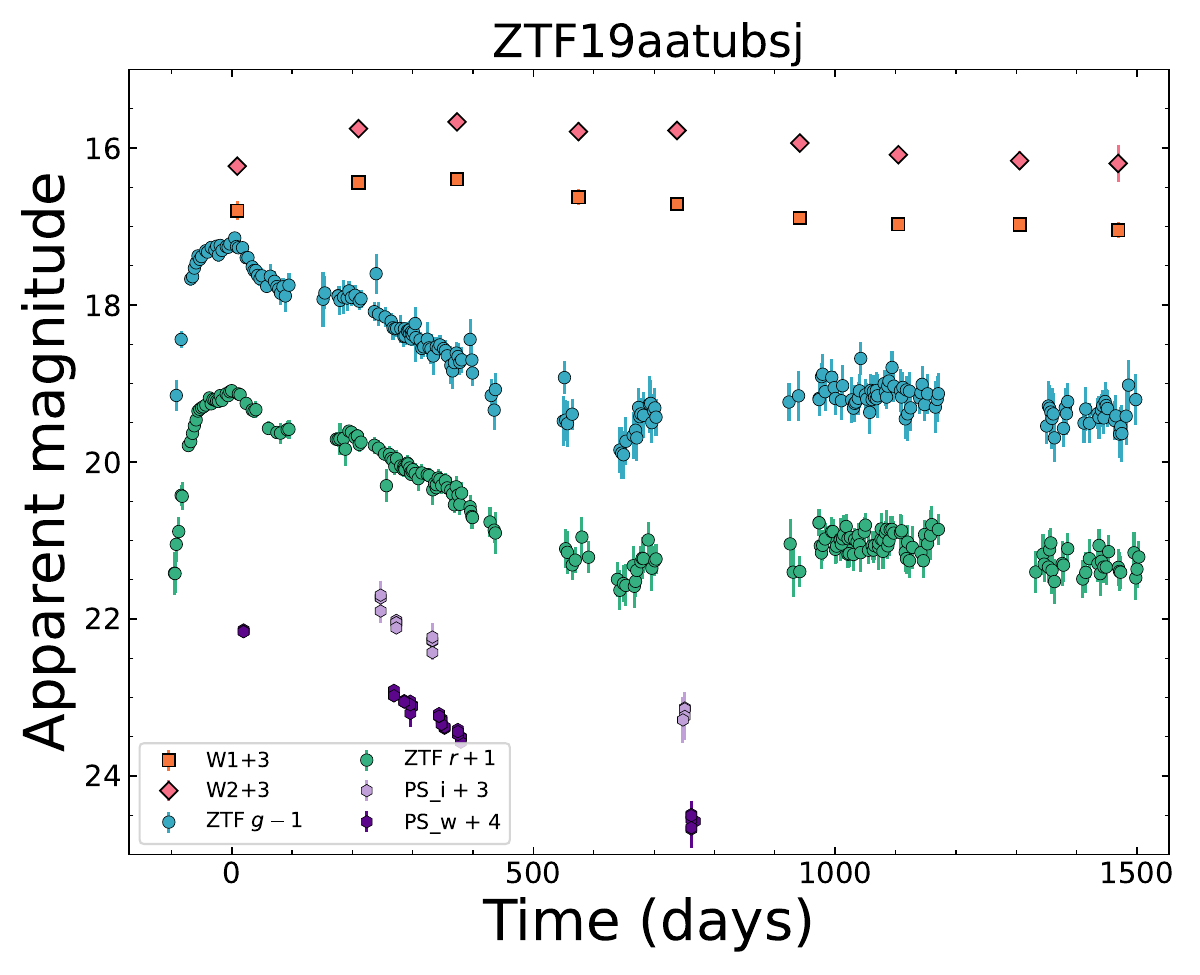} 
     \caption{Observer frame light curve of ZTF19aatubsj relative to $r$-band maximum light.} 
     \label{fig:ZTF19aatubsj_lc_allbands} 
 \end{figure}

\begin{figure} 
     \centering 
     \includegraphics[width=.48\textwidth]{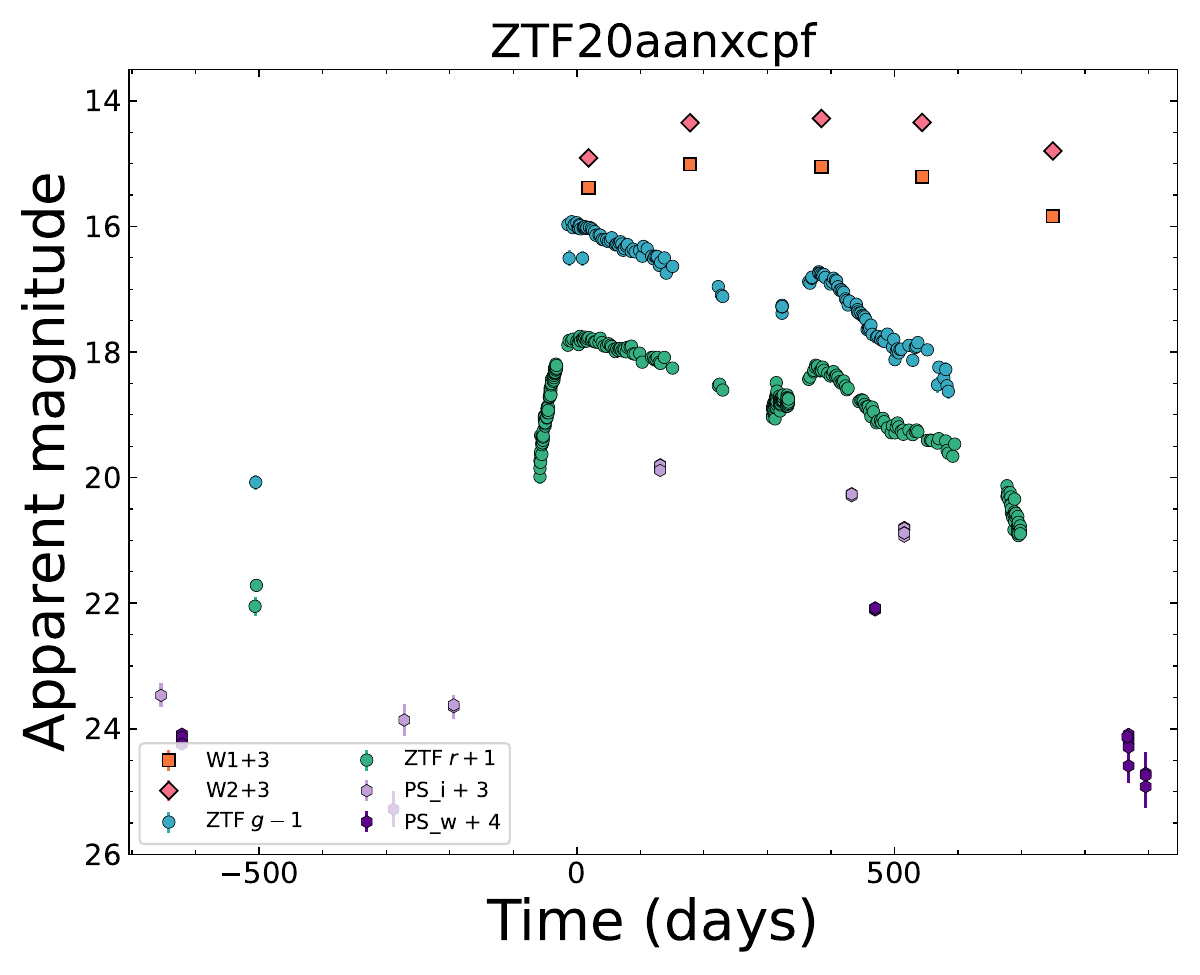} 
     \caption{Observer frame light curve of ZTF20aanxcpf relative to $r$-band maximum light.} 
     \label{fig:ZTF20aanxcpf_lc_allbands} 
 \end{figure}

In this Figs. \ref{fig:ZTF20abrbeie_lc_allbands}-\ref{fig:ZTF20aanxcpf_lc_allbands} we present all the available photometric data for each ANT in our main sample.
\twocolumn
 
%%%%%%%%%%%%%%%%%%%%%%%%%%%%%%%%%%%%%%%%%%%%%%%%%%

% Don't change these lines
\bsp	% typesetting comment
\label{lastpage}
\end{document}